\documentstyle[12pt]{article}
\newcommand{\abs}[1]{\left| #1\right|}
\newcommand{\fnd}[2]{\frac{\textstyle #1}{\textstyle #2}}
\newcommand{\fsml}[2]{\mbox{$\frac{#1}{#2}$}}
\newcommand{\x}[1]{{\textstyle #1}}
\newcommand{\xrm}[1]{{\textstyle \mbox{\rm #1}}}
\newcommand{\bm}[1]{\mbox{\boldmath $#1$}}
\newcommand{\ket}[1]{\mbox{$\left| #1\right\rangle$}}
\newcommand{\bracket}[2]{\mbox{$\left\langle #1\left| #2\right.\right\rangle$}}
\newcommand{\valexp}[1]{\mbox{$\left\langle #1\right\rangle$}}
\newcommand{\Clebsch}[6]{\mbox{$\left(\begin{array}{rrr}
#1 & #2 & #3 \\ [.1cm] #4 & #5 & #6 \end{array}\right)$}}
\begin{document} \baselineskip .7cm
\title{Flavour symmetry of mesonic decay couplings}
\author{
Eef van Beveren\\
{\normalsize\it Departamento de F\'{\i}sica, Universidade de Coimbra}\\
{\normalsize\it P-3000 Coimbra, Portugal}\\
{\small eef@malaposta.fis.uc.pt}\\ [.3cm]
\and
George Rupp\\
{\normalsize\it Centro de F\'{\i}sica das Interac\c{c}\~{o}es Fundamentais}\\
{\normalsize\it Instituto Superior T\'{e}cnico, Edif\'{\i}cio Ci\^{e}ncia}\\
{\normalsize\it P-1096 Lisboa Codex, Portugal}\\
{\small george@ajax.ist.utl.pt}\\ [.3cm]
{\small PACS number(s): 14.40.Cs, 12.39.Pn, 13.75.Lb}\\ [.3cm]
{\small hep-ph/9806248}
}
\date{\today}
\maketitle
\begin{abstract}
We present flavour-symmetric results for the couplings of
quark-antiquark systems to meson-meson channels in the harmonic-oscillator
expansion. We tabulate their values for all possible open and closed
decay channels of pseudo-scalar, vector and scalar mesons. We compare
the predictions of a model that employs these flavour-symmetric
couplings, both with the results of a model which uses explicitly
flavour-dependent couplings, and with experiment. 
\end{abstract}
\clearpage

\section{Introduction}

Particle interactions are described by point-particle vertices in fundamental
theories. Quarks, the basic particles for strong interactions,
are point objects, to our best knowledge, hence assumed to interact
via point-particle vertices in the existing theories:
through a quark-gluon vertex in Quantum Chromodynamics (QCD) \cite{Fritzsch},
through a four-quark vertex in the Nambu-Jona-Lasinio model (NJL)
\cite{Nambu}.

QCD exhibits good agreement with experiment, qualitatively for low and
medium energies, and moreover quantitatively at high energies \cite{QCD},
whereas NJL shows good agreement with experiment only
for energies below 1 GeV \cite{NJL}.
So in the energy interval crucial to meson physics, i.e., ranging
from the two-pion threshold to energies as high as the states in the
bottomonium system, no fundamental theory possesses a satisfactory
descriptive power:
not QCD, because the relevant momentum transfers are too low and thus
the effective color coupling constant is too large for a perturbative
approach, not NJL, because the energies are too high.
Consequently, for a quantitative description of the spectra and scattering of
mesons and baryons, neither of the two theories has sufficient predictive
power for the time being.  Therefore, the use of quark models is still
opportune in this domain of hadronic physics.

Now, ideally a quark model should be derived from QCD, but this is rather
utopian as yet.  As a matter of fact, not even a direct relation between QCD
and confinement has been established and so confinement usually has to be
imposed on the valence quarks of the model \cite{confinement}.
Different models follow distinct strategies to achieve this and the manner in
which confinement is approached distinguishes models among each other.
Moreover, each model has its own very specific purpose, often not
mentioned in too much detail by the authors, which makes it difficult
to compare models. For instance, there exist heavy-quark potential models made
to measure in order to reproduce, with great accuracy, the radial and angular
spectra of charmonium and bottomonium, as well as the electromagnetic
properties of these systems. But if the same potentials are used in the light
quark sector, the results are normally quite bad, especially for radial
excitations, and even possible relativistic corrections are insufficient to
cure the discrepancies. Conversely, sophisticated relativistic models for the
light mesons usually fail to reproduce the correct radial spacings in the
charmonium and bottomonium spectra.

On the other hand, most quark models treat hadrons as manifestly stable bound
states of quarks, simply ignoring the fact that most hadrons are resonances,
some of them even extremely broad, so as to make their very existence
questionable. The standard justification is the conjecture that the effect of
strong decay will be to produce predominantly imaginary mass shifts, thus
allowing to first fit the real parts of the spectra and then to treat the
hadronic widths \em a posteriori, \em with perturbative methods.
However, we know from fundamental principles in scattering theory that real
shifts are generally of the same order as or even larger than the imaginary
ones. Moreover, hadronic loops, i.e., \em virtual \em \/ decay channels, give
rise to \em attractive \em forces, so that the shifts due to, in principle,
all closed decay channels, must be added up so as to produce a negative
mass shift. So not even the true bound states can be treated as pure quark
states. The usual excuse is the unsupported assertion that the effect of closed
channels will be negligible, except near threshold.

However, the Nijmegen unitarised meson model (NUMM), devised
to simultaneously describe meson spectra and meson-meson scattering, from the
light pseudoscalars and vectors \cite{radial}, via the usually awkward scalars
\cite{scalar}, all the way up to the $b\bar{b}$ sector \cite{charm},
showed that both premises are indeed wrong: real shifts are generally
comparable with or larger than the imaginary ones, and the damping of closed
channels is insufficient to make their influence on the ground states
of the spectra negligible. On the contrary, due to the nodal structure
of the radial wave functions and the mentioned additivity property,
the shifts --- real and negative -- of ground states are usually largest
\cite{influence}. Furthermore, no drastic enhancement takes place near
threshold, so that these states cannot be singled out \cite{threshold}.

Having come to the conclusion that, for a truly quantitative description of
mesonic spectra, one must include the coupling to meson-meson channels, the
crucial questions to be raised are how to calculate the involved coupling
constants and which two-meson channels to take along. Here, one should step
back and have another look at the QCD Lagrangian. Realising that, at least
qualitatively, there should be no obvious disagreement between QCD and 
whatever meson model to be used, we are led to respect manifest flavour
blindness. This will impose stringent conditions on how couplings can
be computed and how to select classes of decay channels, since obviously one
cannot take into account an infinite number.

Models which describe the scattering of mesons (and/or baryons) often imitate
the fundamental theories in the sense that interactions take place via
effective point-particle vertices. However, meson (and baryons) are composite
systems, built out of strongly interacting valence quarks, glue, and a
quark-antiquark sea.  So it seems obvious that, when mesons (and baryons) are
considered point objects, some information must get lost.  In this paper we
will demonstrate that this can indeed be the case and how it manifests itself 
in the flavour non-independence of the such-described strong interactions, when
not dealt with carefully.

Point interactions are a powerful tool in constructing theories that not only
consider relativistic kinematics, but also take into account the property of
particle creation and annihilation. However, in applying point interactions
to composite systems, one should include all hidden degrees of freedom. Flavour
is just one such degree of freedom. Angular momentum and spin are others which
should be properly included. At present, it is opportune to model the internal
degrees of freedom and next to integrate them out for the determination of
effective point couplings.  A consistent way of doing so, which moreover
preserves flavour independence, is described below.

The organisation of this paper is as follows. In section \ref{flavour},
we discuss the general philosophy behind a simple model for flavour symmetry.
This model is then exposed in section \ref{flavmod}. The intensities of the
three-meson vertices for meson decay into meson-meson pairs are given in
section \ref{3mesonvertex}. Results are discussed in section
\ref{results}. The consequences of flavour (in)dependence are studied for
two different, though similar, models in section \ref{comparison}.
Some essential formulae are collected in the appendices
\ref{rearrangement} and \ref{scalarmixing}.

\section{Flavour symmetry}
\label{flavour}

Since strong interactions are independent of flavour, the probability to
create a quark-antiquark pair out of the vacuum cannot depend on the flavour
of the quark and the antiquark. However, it obviously depends on the masses
involved. But if for a moment we assume that the flavour masses,
or at least the effective quark masses in the relevant energy interval
of the three lowest flavours, {\it up}, {\it down} and {\it strange}, are
equal, then the corresponding probabilities of pair creation should be equal.
Let us apply this principle to the strong coupling of a meson to a pair of
mesons. Here, we assume that the related strong decay processes are triggered
by the creation of a flavourless quark-antiquark pair.

In order to set the picture, we consider a simple model in which the initial
meson is described by a confined quark-antiquark system of any flavour,
given by

\begin{equation}
a\bar{b}
\;\;\; ,
\label{qqbar}
\end{equation}

\noindent
where $a$ and $b$ represent any of the three flavours under discussion,
and the final pair of decay products is described by a system of two
freely moving mesons, which represent any of the three combinations

\begin{equation}
(a\bar{u})\; +\; (u\bar{b})\;\; ,\;\;
(a\bar{d})\; +\; (d\bar{b})\;\; ,\;\;
(a\bar{s})\; +\; (s\bar{b})\;\; .
\label{mesonmeson}
\end{equation}

Also, let us, for a moment, assume that no further quantum numbers are
involved. Then, flavour symmetry demands that the probabilities are equal for
the $a\bar{b}$ system to decay into any of the three channels of formula
(\ref{mesonmeson}). In particular, when under full flavour symmetry
the six mesons represented in formula (\ref{mesonmeson}) have all the
same mass, then the experimental results for the decay of system
(\ref{qqbar}) into any of the three channels of formula (\ref{mesonmeson})
should be indistinguishable.

Let the three decay coupling constants of the process under consideration
be represented by, respectively,

\begin{equation}
g(a,b;u)\;\; ,\;\;
g(a,b;d)\;\; ,\;\;
g(a,b;s)\;\; .
\label{koppeling}
\end{equation}

\noindent
Then, assuming full flavour symmetry, we have the identities

\begin{equation}
\left\{ g(a,b;u)\right\}^{2}\; =\;
\left\{ g(a,b;d)\right\}^{2}\; =\;
\left\{ g(a,b;s)\right\}^{2}
\;\;\; ,
\label{flasym1}
\end{equation}

\noindent
and, moreover, for the total decay intensity $\Gamma (a,b)$ the relation

\begin{equation}
\Gamma (a,b)\; =\; A\;\left[\left\{ g(a,b;u)\right\}^{2}\; +\;
\left\{ g(a,b;d)\right\}^{2}\; +\;
\left\{ g(a,b;s)\right\}^{2}\right]
\;\;\; ,
\label{flasym2}
\end{equation}

\noindent
where the proportionality factor $A$ is also completely flavour symmetric,
which means constant here.
Furthermore, one has that, under flavour symmetry, $\Gamma (a,b)$ must be
independent of the flavours $a$ and $b$.

Unfortunately, quarks are fermions and mesons are spatially extended systems,
and hence spin and spatial quantum numbers do play an important r\^{o}le in
the decay of a meson into a pair of mesons.
Nevertheless, it remains possible to construct coupling constants which
have the property that the total decay probability is independent of the
flavour of the decaying meson in the limit of equal masses, as we will see
below.
\clearpage

\section{Modelling full flavour symmetry}
\label{flavmod}

When a normalised wave function $\psi$ is expanded on a complete orthonormal
basis $\phi_{n}$, for $n=0,1,2,\dots$, according to

\begin{equation}
\psi\; =\;\sum_{n=0}^{\infty}\; c_{n}\phi_{n}
\;\;\; ,
\label{expand}
\end{equation}

\noindent
one has for the expansion coefficients $c_{n}$ the property

\begin{equation}
\sum_{n=0}^{\infty}\; \abs{c_{n}}^{2}\; =\; 1
\;\;\; .
\label{complete}
\end{equation}

\noindent
It is exactly property (\ref{complete}) that leads to flavour symmetry.

Let us consider a system of two quarks and two antiquarks, like any
of the three combinations of formula (\ref{mesonmeson}). One complete basis
for the Hilbert space of such a system can be constructed by taking products
of the internal wave function of the $a\bar{b}$ system, specifying
thereby its internal spatial and flavour quantum numbers, the internal
wave function of the $q\bar{q}$ (either $u\bar{u}$, $d\bar{d}$ or
$s\bar{s}$) system, and the relative wave function of the two
subsystems. Another complete basis for this Hilbert space consists of
products of the internal wave function of the $a\bar{q}$ system, the
internal wave function of the $q\bar{b}$ system, and the relative wave
function of those two subsystems. Any wave function describing one of the
two-quark-two-antiquark systems (\ref{mesonmeson}) can be expanded in
either of the above-defined bases.

Such an expansion takes a particularly manageable form if the four partons
are supposed to move in a harmonic oscillator potential with universal
frequency. In that case the spatial quantum numbers are linearly related
to the total energy of the system, which gives rise to finite bases at
each energy level and hence to finite expansions. The flavour-symmetry
condition (\ref{complete}) then becomes a finite sum, which makes it easy for
verification. Furthermore, the restriction to harmonic oscillators is not a
real limitation, since any other basis can always be expanded in the
corresponding harmonic oscillator basis, $\{ n\}$, according to

\begin{equation}
\left\langle M_{1}M_{2}\left| V\right| M\right\rangle\; =\;
\sum_{\{ n,n'\}}
\left\langle M_{1}M_{2}\left| n'\right.\right\rangle
\left\langle n' \left| V\right| n\right\rangle
\left\langle n\left| M\right.\right\rangle
\;\;\; .
\label{anypot}
\end{equation}

\noindent
Here, $V$ represents the interaction Hamiltonian which describes the
transitions between the quark-antiquark system $a\bar{b}$ and the
two-meson channels. We assume that the spatial, or momentum-dependent,
part of the matrix elements $\left\langle n' \left| V\right| n\right\rangle$
is flavour independent and that the flavour-dependent parts are constants.

The expansion of a particular many-particle wave function into a
specific basis for well-defined subsystems, or {\it recoupling},
has been studied a great deal in the past. The related coefficients
for the harmonic oscillator basis are known as {\it Moshinsky brackets}.
Moshinsky brackets are well-known coefficients of recoupling in Nuclear
Physics; see Ref.~\cite{Talmi} for their definition.
The group-theoretical implications of parton
recoupling in the harmonic-oscillator approximation have been studied
exhaustively in Ref.~\cite{Barg60}.
Their application to meson decay has for the first time been formulated in
Ref.~\cite{Ribe82}. A full generalisation for the spatial part of the
recoupling constants, which includes all possible quantum numbers for any
number of (bosonic) partons, can be found in Ref.~\cite{rekpplng}.
The inclusion of fermionic and flavour degrees of freedom, which leads to an
analytic expression for the coupling constants of any meson to any of its
two-meson real or virtual decay channels, is given in Ref.~\cite{kpplng},
for the case that the new valence pair is created with $^{3}P_{0}$
quantum numbers. A Fortran source program is available on request.
\clearpage

\section{Coupling constants for three-meson vertices}
\label{3mesonvertex}

Within the above-outlined formalism, we assume that mesons can be
classified by the quantum numbers of their valence constituent
quark-antiquark distributions, i.e.,

\begin{equation}
\xrm{meson}\left( j,M,\ell ,s,n,{\cal M}\right)
\;\;\; .
\label{meson}
\end{equation}

\noindent
The quantum numbers $j$ and $M$ in formula (\ref{meson}) represent,
respectively, the spin of the meson and its $z$-component.
Alternatively, $j$ represents the total angular momentum of the relative
motion in the quark+antiquark system which describes the meson.
The quantum numbers $\ell$, $s$ and $n$ stand, respectively, for
the orbital angular momentum, the spin, and the radial excitation of the
constituents of the meson. Finally, $\cal M$ represents the $3\times 3$ flavour
matrix which indicates the valencies of the quark and the antiquark.

Here, we study the decay intensities for the following processes:

\begin{equation}
\xrm{meson}\left( J,J_{z},\ell ,s,n,{\cal M}_{C}\right)
\;\longrightarrow\;
\xrm{meson}\left( j_{1},M_{1},\ell_{1},s_{1},n_{1},{\cal M}_{A}\right) +
\xrm{meson}\left( j_{2},M_{2},\ell_{2},s_{2},n_{2},{\cal M}_{B}\right)
\; .
\label{MMdecay}
\end{equation}

\noindent
This is not a completely satisfactory notation, since the spin $z$-components
$M_{1}$ and $M_{2}$ of the decay products in formula (\ref{MMdecay}) are not
supposed to be observable and, moreover, the quantum numbers which
characterise the relative motion of the decay products are not specified in
formula (\ref{MMdecay}), despite being equally important.
Let us indicate the orbital quantum numbers of the two-meson system by means
of an index, $r$, and hence denote
the orbital angular momentum of the two mesons by $\ell_{r}$,
the total spin by $s_{r}$, and the radial excitation of the relative motion in
the two-meson system by $n_{r}$.
The total angular momentum of the two-meson system and its $z$-component are,
due to angular-momentum conservation, given by $J$ and $J_{z}$, respectively.

The decay probability for the process (\ref{MMdecay}) is then, following
the formalism developed in Ref.~\cite{kpplng}, given by the following matrix
element:

\begin{eqnarray}
 & & \;\;\;\;\;\bracket{J,J_{z},j_{1},\ell_{1},s_{1},n_{1},
j_{2},\ell_{2},s_{2},n_{2},\ell_{r},s_{r},n_{r},A,B}
{J,J_{z},\ell ,s,n,C}\; =
\label{decint} \\ [.3cm] & &
=\;\xrm{Tr}\left\{\;
{\cal M}_{A}{\cal M}_{B}{{\cal M}_{C}}^{T}
\;\bracket{J,J_{z},j_{1},\ell_{1},s_{1},n_{1},
j_{2},\ell_{2},s_{2},n_{2},\ell_{r},s_{r},n_{r}}
{J,J_{z},\ell ,s,n,\bm{\alpha}_{ABC}}
\right.\; +
\nonumber \\ [.3cm] & &
+\left.\;
{\cal M}_{B}{\cal M}_{A}{{\cal M}_{C}}^{T}
\;\bracket{J,J_{z},j_{1},\ell_{1},s_{1},n_{1},
j_{2},\ell_{2},s_{2},n_{2},\ell_{r},s_{r},n_{r}}
{J,J_{z},\ell ,s,n,\bm{\alpha}_{BAC}}\;
\right\}\;\;\; .
\nonumber
\end{eqnarray}

\noindent
The spatial parts in each of the two terms of the transition element
(\ref{decint}) are denoted by

\begin{equation}
\bracket{J,J_{z},j_{1},\ell_{1},s_{1},n_{1},
j_{2},\ell_{2},s_{2},n_{2},\ell_{r},s_{r},n_{r}}
{J,J_{z},\ell ,s,n,\bm{\alpha}}
\;\;\; ,
\label{spatial}
\end{equation}

\noindent
and are defined and explained in Appendix~\ref{rearrangement}.

Since for many purposes it is sufficient to have flavour-symmetric
coupling constants for the corresponding strong decay channels of pseudoscalar,
vector, and scalar mesons, we tabulate the probabilities for the three-meson
vertices of those decay processes; 
for a pseudoscalar meson in Table \ref{pseudoscalar}, for a vector meson
in Table \ref{vector}, and for a scalar meson in Table \ref{scalar}. 
In order to maintain the tables as condensed as possible, we represent
mesons by symbols and by their quantum numbers.
Since we assume that isospin is indeed a perfect symmetry, we may
represent all members of an isomultiplet by the same symbol,
for which we just have chosen the letters and numbers $t$, $d$, $8$ and $1$,
according to the identification given in Table \ref{partid}.
Now let us just analyse one horizontal line of one of the three tables,
to make sure that the reader understands what the numbers represent. Let us
take the fourth line of Table \ref{pseudoscalar}. In the first column we
find four zeroes, representing the internal spatial quantum numbers $j$,
$\ell$, $s$, and $n$ of the first decay product, $M_{1}$, which hence
characterises a meson out of the lowest-lying ($n=0$) pseudoscalar nonet.
In the second column we find similarly that the second decay product,
$M_{2}$, represents a meson out of the lowest-lying vector nonet.
In the third column we find the quantum numbers for the relative motion of
$M_{1}$ and $M_{2}$, i.e., $P$-wave ($\ell_{r}=1$) with total spin one
($s_{r}=1$) in the lowest radial excitation ($n_{r}=0$). Since the table
refers to the real or virtual decays of the lowest-lying pseudoscalar meson
nonet ($J\ell sn=0000$, indicated in the top of the table),
the next four columns refer to
its isotriplet member, which is the pion. We then find that the pion couples
with a strength $\sqrt{1/6}$ to the $tt$ (isotriplet-isotriplet) channel,
which, following Table \ref{partid} and the above-discussed particle
assignments to $M_{1}$ and $M_{2}$, i.e., pseudoscalar and vector
respectively, represents in this case the $\pi\rho$ channel. Following a
similar reasoning, we find that the pion couples
with a strength $\sqrt{1/12}$ to $KK^{\ast}$.
The total coupling of a pion to pseudoscalar-vector channels is given in the
column under $T$ by $\sqrt{1/4}$, which is the square root of the quadratic
sum of the two previous couplings, i.e., $\sqrt{1/6+1/12}$.

The next set of coupling constants refer to the real or virtual (actually only
virtual) decays of a kaon. We find
$\sqrt{1/8}$ to $td$, which represents both of the possibilities
pseudoscalar (isotriplet) + vector (isodoublet), i.e., $\pi K^{\ast}$, and
pseudoscalar (isodoublet) + vector (isotriplet), i.e., $K\rho$,
each with half of the intensity that is given in the table, and therefore
one has for the kaon the coupling constants
$\sqrt{1/16}$ to $\pi K^{\ast}$ and $\sqrt{1/16}$ to $K\rho$. 
Next, we find in the table that the kaon couples with $\sqrt{1/8}$ to $d8$,
which represents both of the possibilities,
pseudoscalar (isodoublet) + vector ($SU_{3}$-octet isoscalar),
i.e., $K$ + some admixture of the $\omega$ and $\phi$ mesons, and
pseudoscalar ($SU_{3}$-octet isoscalar) + vector (isodoublet),
i.e., some admixture of $\eta$'s + $K^{\ast}$,
each with half of the intensity that is given in the table, and hence one
extracts for the kaon the coupling constants
$\sqrt{1/16}$ to $K+(\omega ,\phi )$ and
$\sqrt{1/16}$ to $(\eta ,\eta ')+K^{\ast}$.
The kaon does not couple to the $d1$ channels in pseudoscalar + vector,
which represent the channels with one isodoublet and one $SU_{3}$-singlet.
The total coupling for the kaon to its pseudoscalar + vector decay channels
sums up to $\sqrt{1/4}$, as one verifies in the column under $T$.
The next two sets of coupling constants refer similarly to the decay
modes of the isoscalar, either $SU_{3}$-octet or $SU_{3}$-singlet, partners
of the pseudoscalar nonet. Mixings can be done by hand as examplified
in Appendix~\ref{scalarmixing}.

Notice that for the $SU_{3}$-octet members one has flavour symmetry for
each horizontal line in the tables.
This does not go through for the $SU_{3}$-singlet partners, with the
exception of scalar meson decay (Table \ref{scalar}), where all horizontal
lines have the same total coupling in each subsection of the table.
However, all columns under $T$ sum up to 1, representing full flavour symmetry
once all possible decay channels have been accounted for.

\section{Results}
\label{results}

The pole structures of the scattering matrices for $P$-wave meson-meson
scattering, or equivalently, the radial spectra of heavy and light
pseudoscalar and vector mesons, were studied in the, largely non-relativistic,
coupled-channel NUMM, published in Ref.~\cite{radial} hereafter referred to as
B83, in which the authors parametrised confinement by a universal frequency,
the same for all flavours, including charm and bottom.
The universal frequency and a flavour-independent overall coupling constant,
representing the probability for the creation of a $^{3}P_{0}$ light
quark-antiquark pair, were sufficient to obtain theoretical predictions
for phase shifts and scattering cross sections, or equivalently,
for central resonance positions and widths, which were in reasonable agreement
with the data.
All relative couplings were exactly taken as given in Tables
\ref{pseudoscalar} and \ref{vector}, though some of these had been derived in a
more empiric way, and then extended in order to also include the heavy-quark
systems.  This extension is quite trivial and will not be discussed here.
The only flavour non-invariance came from the quark masses and the two-meson
thresholds, all other ingredients were the same for all flavours. Of course,
many of the decay channels were omitted, assuming their thresholds  to be high
enough in energy, so as not to have too much importance for the details of the
scattering processes at much lower energies. But this is only a practical
ingredient, not to be confused with flavour breaking.

In Ref.~\cite{EMtrans}, the electromagnetic transitions in the charmonium and
bottomonium systems were studied, using the quark and meson distributions
from B83, with good results, indicating that not only
the pole structures of the scattering matrices, but also the related
wave functions stood the confrontation with experiment.

In Ref.~\cite{scalar}, hereafter referred to as B86, the pole structure of the
scattering matrix was inspected for $S$-wave meson-meson scattering.  Since the
model was the same as for Ref.~\cite{radial}, using exactly the same
universal frequency, flavour-independent overall coupling constant, and quark 
masses, the calculated phase shifts and scattering cross sections could be 
considered genuine theoretical predictions.
The agreement with the data was unexpected, especially because it had not been
the objective of the model, neither was the model constructed towards
fitting the $S$-wave scattering data.
All relative couplings were exactly taken as given in Table \ref{scalar}.
Also here, only those two-meson channels were taken into account which
contain members of the lowest-lying pseudoscalar and vector nonets.
That such a procedure does not break flavour invariance may be explicitly
verified by checking the first, third, and last line in Table \ref{scalar}.

The observed flavour independence of strong interactions is a very important
ingredient for low-energy hadron physics and woven into the NUMM;
first, by the universal frequency, which makes the ratio
of the kinetic term and the potential term of the model flavour independent
and hence also the level splittings; second, by the intensities of the
three-meson vertices for the coupling to the various decay channels.
\clearpage

\section{Comparison of two models}
\label{comparison}

As stated in the introduction, it is not easy to compare meson models,
but here we will pay attention to the comparison of the NUMM
with a model \cite{Toern95}, hereafter referred to as T95, which
is tailor-made for scalar mesons or, in other words, for $S$-wave meson-meson
scattering. The latter model, a revised version of the Helsinki unitarised
quark model (HUQM), was confronted with experiment in an analysis published in
Ref.~\cite{Toern96}, hereafter referred to as TR96.
Based on the good agreement of its theoretical predictions with the available
experimental phase shifts, one may be inclined to accept all further
conclusions presented in the same publication, such as the existence and
location of resonances. However, the authors failed to find the complex-energy
pole corresponding to the established $f_0$(1500) resonance. Furthermore, they
also did not find a light $K_0^*$, i.e., the old $\kappa$, somewhere between
700 and 1100 MeV. Although the latter resonance is not (yet) established
experimentally, it has recently received renewed phenomenological and
theoretical support \cite{Ish97,Sch98,Rij98} (see also Ref.~\cite{comment}).
Moreover, its absence in Nature would imply a breaking of the conventional
nonet pattern for mesons. On the other hand, if a light $K_0^*$ is confirmed,
then there exist unmistakable experimental candidates for \em two \em \/
complete scalar nonets, as predicted by B86.
So it is intriguing to figure out why model T95/TR96, which is very
similar in its philosophy and also in observing a resonance doubling, at least
for some states, does not reproduce this resonance.

In Table \ref{tabspsps}, we collect the intensities for strong scalar-meson
decay into a pair of pseudoscalar mesons, under the assumption that pions,
kaons, and eta's have equal masses, as given by models B86 and T95/TR96.
For the purpose of comparison, we have multiplied the values given in
B86 by a constant factor. The resulting values can also be
read from the first line of Table \ref{scalar} when renormalised
(i.e., multiplied by a factor 24), and when isoscalar mixing has been
dealt with as outlined in Appendix~\ref{scalarmixing}.
Now notice that, by using formula (\ref{flasym2}), the total decay intensities
stemming from Ref.~\cite{scalar} become equal to $A$ for all scalar
mesons, as demanded by flavour symmetry, i.e.,

\begin{equation}
\Gamma\left( a_{0}\right)\; =\;
\Gamma\left( \kappa\right)\; =\;
\Gamma\left( f_{0}\; ,n\bar{n}\right)\; =\;
\Gamma\left( f_{0}\; ,s\bar{s}\right)\; =\; A
\;\;\; .
\label{flasym3}
\end{equation}

For model T95/TR96, the comparable intensities are derived from
a point-particle approach to the three-meson vertex, which results in
coupling constants given by

\begin{equation}
\xrm{Tr}\left({\cal M}_{A}{\cal M}_{B}{\cal M}_{C}\right)
\;\;\; ,
\label{simple}
\end{equation}

\noindent
where $A$, $B$ and $C$ stand for the three mesons involved at the vertex
$C\rightarrow AB$, and ${\cal M}_{X}$ is the $3\times 3$ flavour matrix for
meson $X$.
Flavour symmetry as from Eq.~(\ref{simple}), and using formula
(\ref{flasym2}), yields in this case the flavour-dependent result

\begin{equation}
\Gamma\left( a_{0}\right)\; =\;
\Gamma\left(\kappa\right)\; =\;
\fnd{3}{5}\Gamma\left( f_{0}\; n\bar{n}\right)\; =\;
\fnd{3}{4}\Gamma\left( f_{0}\; s\bar{s}\right)\; =\; A
\;\;\; ,
\label{flasym4}
\end{equation}

\noindent
in contrast with the results shown in Eq.~(\ref{flasym3}).

It might be surprising that such a seemingly flavour-symmetric
vertex leads to flavour {\em non}-independence when applied to
strong meson decay. However, if one does not take into account the internal
structure of the various quark-antiquark systems involved, then
normalisations, so essential to wave functions or distributions, are swept
under the rug, and thus the above vertex intensity (\ref{simple})
leads to flavour-dependent results.
It appears to be for this reason that the authors of TR96, which base their
calculations on the coupling constants from T95, do not observe any resonance
doubling for the isodoublet and one of the two isoscalars, and therefore miss
the $K^{\ast}_{0}$(700-1100) and $f_0$(1500) poles needed to complete two
scalar meson nonets.

The normalisation factors that are relevant to formula (\ref{flasym3})
are given in formulae (\ref{decABC}) and (\ref{normtab}) of Appendix
\ref{rearrangement}. In Appendix~\ref{scalarmixing}, we show how they
lead exactly to the factors $\fsml{3}{5}$ and $\fsml{3}{4}$ which
are necessary to compensate the flavour-dependence of formula (\ref{flasym4}).
\clearpage

\appendix

\section{Rearrangement coefficients}
\label{rearrangement}

The spatial parts of the matrix elements (\ref{decint}) are,
following the formalism developed in Ref.~\cite{kpplng},
in the approximation of equal flavour masses just given by 
Clebsch-Gordonary and some overlap integrals, amounting to

\begin{eqnarray}
 & & \;\;\;\;\;\left.\begin{array}{l}
\bracket{J,J_{z},j_{1},\ell_{1},s_{1},n_{1},
j_{2},\ell_{2},s_{2},n_{2},\ell_{r},s_{r},n_{r}}
{J,J_{z},\ell ,s,n,\bm{\alpha}_{ABC}}\\ [.5cm]
\bracket{J,J_{z},j_{1},\ell_{1},s_{1},n_{1},
j_{2},\ell_{2},s_{2},n_{2},\ell_{r},s_{r},n_{r}}
{J,J_{z},\ell ,s,n,\bm{\alpha}_{BAC}}\end{array}\right\}\; =
\nonumber \\ [.5cm] & &
=\;\fnd{1}{\sqrt{1+
\bracket{C}{SU(3)_\xrm{\scriptsize flavour}\xrm{-singlet}}
\delta (^{2s+1}\ell_{J},^{3}P_{0})\delta_{n0}}}\;\;
\sum_\x{\{\mu\} ,\{ m\} ,\{ M\}}
\label{decABC} \\ [.3cm] & &
\Clebsch{s_{r}}{\ell_{r}}{J}{M_{r}}{m_{r}}{J_{z}}
\Clebsch{j_{1}}{j_{2}}{s_{r}}{M_{1}}{M_{2}}{M_{r}}
\Clebsch{\ell_{1}}{s_{1}}{j_{1}}{m_{1}}{\mu_{1}}{M_{1}}
\Clebsch{\ell_{2}}{s_{2}}{j_{2}}{m_{2}}{\mu_{2}}{M_{2}}\;\times
\nonumber \\ [.3cm] & & \times\;
\Clebsch{\ell}{s}{J}{m_{\ell}}{\mu_{s}}{J_{z}}
\Clebsch{1}{1}{0}{m}{-m}{0}
\Clebsch{\fsml{1}{2}}{\fsml{1}{2}}{s_{1}}{\mu_{a}}{\mu_{b}}{\mu_{1}}
\Clebsch{\fsml{1}{2}}{\fsml{1}{2}}{s_{2}}{\mu_{c}}{\mu_{d}}{\mu_{2}}\;\times
\nonumber \\ [.3cm] & & \times\;
\left\{\begin{array}{l}
\Clebsch{\fsml{1}{2}}{\fsml{1}{2}}{s}{\mu_{a}}{\mu_{d}}{\mu_{s}}
\Clebsch{\fsml{1}{2}}{\fsml{1}{2}}{1}{\mu_{c}}{\mu_{b}}{-m}
\left(\begin{array}{ccccccc}
n & \ell & m_{\ell} &  & n_{1} & \ell_{1} & m_{1}\\ [.1cm]
0 & 1 & m &  & n_{2} & \ell_{2} & m_{2}\\ [.1cm]
0 & 0 & 0 &  & n_{r} & \ell_{r} & m_{r}
\end{array}\right)_\x{\bm{\alpha}_{ABC}}\\ [.6cm]
\Clebsch{\fsml{1}{2}}{\fsml{1}{2}}{s}{\mu_{c}}{\mu_{b}}{\mu_{s}}
\Clebsch{\fsml{1}{2}}{\fsml{1}{2}}{1}{\mu_{a}}{\mu_{d}}{-m}
\left(\begin{array}{ccccccc}
n & \ell & m_{\ell} &  & n_{1} & \ell_{1} & m_{1}\\ [.1cm]
0 & 1 & m &  & n_{2} & \ell_{2} & m_{2}\\ [.1cm]
0 & 0 & 0 &  & n_{r} & \ell_{r} & m_{r}
\end{array}\right)_\x{\bm{\alpha}_{BAC}}\end{array}\right.
\nonumber
\end{eqnarray}

\noindent
where the sum is over all $\mu$'s, $m$'s, and $M$'s that appear in the formula,
where

\begin{eqnarray}
\bracket{C}{SU(3)_\xrm{\scriptsize flavour}\xrm{-singlet}}\; = & &
\nonumber \\ [.3cm]
0 & \xrm{for} & \ket{C}\;\;\;\xrm{orthogonal to the }
SU(3)_\xrm{\scriptsize flavour}\xrm{-singlet state}
\;\;\; ,\nonumber \\ [.3cm]
\fsml{1}{3} & \xrm{for} & \ket{C}\; =\;
\ket{u\bar{u}}\; ,\;\;\;
\ket{d\bar{d}}\; ,\;\;\xrm{or}\;\;\;
\ket{s\bar{s}}
\;\;\; ,\nonumber \\ [.3cm]
\fsml{2}{3} & \xrm{for} & \ket{C}\; =\;\sqrt{\fsml{1}{2}}\left\{
\ket{u\bar{u}}+\ket{d\bar{d}}\right\}
\;\;\; ,\label{normtab} \\ [.3cm]
1 & \xrm{for} & \ket{C}\; =\;\sqrt{\fsml{1}{3}}\left\{
\ket{u\bar{u}}+\ket{d\bar{d}}+\ket{s\bar{s}}\right\}
\nonumber
\end{eqnarray}

\noindent
and where

\begin{equation}
\delta (^{2s+1}\ell_{J},^{3}P_{0})\; =\;
\delta_{J0}\delta_{\ell 1}\delta_{s1}
\;\;\; .
\label{delta3P0}
\end{equation}

The central part of formula~(\ref{decABC}) is constituted by
the rearrangement coefficients,
which can be given by the following diagramatic representation

\begin{equation}
\left(\begin{array}{ccccccc}
n & \ell & m_{\ell} &  & n_{1} & \ell_{1} & m_{1}\\ [.1cm]
0 & 1 & m &  & n_{2} & \ell_{2} & m_{2}\\ [.1cm]
0 & 0 & 0 &  & n_{r} & \ell_{r} & m_{r}
\end{array}\right)_\x{\bm{\alpha}}\; =\;\;\;\;
\begin{picture}(240,80)(0,60)
\linethickness{2pt}
\multiput(0,0)(0,60){3}{\line(1,0){60}}
\multiput(180,0)(0,60){3}{\line(1,0){60}}
\thinlines
\put(60,0){\line(1,0){50}}
\put(120,0){\makebox(0,0){$\alpha_{33}$}}
\put(130,0){\line(1,0){50}}
\put(60,0){\line(2,1){35}}
\put(100,20){\makebox(0,0){$\alpha_{23}$}}
\put(105,22.5){\line(2,1){75}}
\put(60,0){\line(1,1){15}}
\put(80,20){\makebox(0,0){$\alpha_{13}$}}
\put(85,25){\line(1,1){95}}
\put(60,60){\line(2,-1){71}}
\put(140,20){\makebox(0,0){$\alpha_{32}$}}
\put(149,15.5){\line(2,-1){31}}
\put(60,60){\line(1,0){20}}
\put(90,60){\makebox(0,0){$\alpha_{22}$}}
\put(100,60){\line(1,0){80}}
\put(60,60){\line(2,1){75}}
\put(140,100){\makebox(0,0){$\alpha_{12}$}}
\put(145,102.5){\line(2,1){35}}
\put(60,120){\line(1,0){50}}
\put(120,120){\makebox(0,0){$\alpha_{11}$}}
\put(130,120){\line(1,0){50}}
\put(60,120){\line(2,-1){31}}
\put(100,100){\makebox(0,0){$\alpha_{21}$}}
\put(109,95.5){\line(2,-1){71}}
\put(60,120){\line(1,-1){15}}
\put(80,100){\makebox(0,0){$\alpha_{31}$}}
\put(85,95){\line(1,-1){95}}
\put(30,5){\makebox(0,0)[cb]{$0,0,0$}}
\put(30,65){\makebox(0,0)[cb]{$0,1,m$}}
\put(30,125){\makebox(0,0)[cb]{$n,\ell ,m_{\ell}$}}
\put(210,5){\makebox(0,0)[cb]{$n_{r},\ell_{r},m_{r}$}}
\put(210,65){\makebox(0,0)[cb]{$n_{2},\ell_{2},m_{2}$}}
\put(210,125){\makebox(0,0)[cb]{$n_{1},\ell_{1},m_{1}$}}
\end{picture}
\label{reardiagram}
\end{equation}
\vspace{2cm}

\noindent
The upper-left external line carries the relevant quantum numbers of
the initial meson in formula (\ref{MMdecay}), the middle-left external line
the quantum numbers of the $^{3}P_{0}$ $q\bar{q}$-pair, and the lower-left
external line the quantum numbers of the relative motion of the two
quark-antiquark systems, which, to lowest order, is supposed to be in its
ground state. The external lines on the right-hand side of the diagram carry
the quantum numbers of the decay products of formula (\ref{MMdecay}) and their
relative motion.

As explained in Ref.~\cite{rekpplng}, each of the internal lines ${ij}$ of
diagram (\ref{reardiagram}) carries the set of quantum numbers
$\left\{ n_{ij},\ell_{ij},m_{ij}\right\}$, over all possibilities of which
must be summed, thereby respecting partial quantum-number conservation at
each vertex, i.e.,

\begin{eqnarray}
 & & \;\;\;\;\;
\left(\begin{array}{ccccccc}
n & \ell & m_{\ell} &  & n_{1} & \ell_{1} & m_{1}\\ [.1cm]
0 & 1 & m &  & n_{2} & \ell_{2} & m_{2}\\ [.1cm]
0 & 0 & 0 &  & n_{r} & \ell_{r} & m_{r}
\end{array}\right)_\x{\bm{\alpha}}
\; =\;
(-1)^\x{n+n_{1}+n_{2}+n_{r}}\;\left(\fnd{\pi}{4}\right)^{3}\;
\sqrt{\left( n!\; n_{1}!\; n_{2}!\; n_{r}!\right)}
\nonumber \\ [.3cm] & &
\sqrt{\left(\fnd{
\Gamma\left( 2n+\ell+\fsml{3}{2}\right)
\Gamma\left(\fsml{5}{2}\right)
\Gamma\left(\fsml{3}{2}\right)
\Gamma\left( 2n_{1}+\ell_{1}+\fsml{3}{2}\right)
\Gamma\left( 2n_{2}+\ell_{2}+\fsml{3}{2}\right)
\Gamma\left( 2n_{r}+\ell_{r}+\fsml{3}{2}\right)}
{\left( 2\ell +1\right)
\left( 3\right)
\left( 2\ell_{1}+1\right)
\left( 2\ell_{2}+1\right)
\left( 2\ell_{r}+1\right)}\right)}
\nonumber \\ [.3cm] & &
\sum_\x{\left\{ n_{ij},\ell_{ij},m_{ij}\right\}}\;\;
\prod_\x{i,j}\;\; \left(\alpha_{ij}\right)^\x{2n_{ij}+\ell_{ij}}
\fnd{\left( 2\ell_{ij}+1\right)}
{n_{ij}!\;\Gamma\left( n_{ij}+\ell_{ij}+\fsml{3}{2}\right)}
\nonumber \\ [.3cm] & &
\delta\left( 2\left[ n_{11}+n_{21}+n_{31}\right] +
\ell_{11}+\ell_{21}+\ell_{31}\; ,\; 2n+\ell\right)
\nonumber \\ [.3cm] & &
\delta\left( 2\left[ n_{12}+n_{22}+n_{32}\right] +
\ell_{12}+\ell_{22}+\ell_{32}\; ,\; 1\right)
\nonumber \\ [.3cm] & &
\delta\left( 2\left[ n_{13}+n_{23}+n_{33}\right] +
\ell_{13}+\ell_{23}+\ell_{33}\; ,\; 0\right)
\nonumber \\ [.3cm] & &
\delta\left( 2\left[ n_{11}+n_{12}+n_{13}\right] +
\ell_{11}+\ell_{12}+\ell_{13}\; ,\; 2n_{1}+\ell_{1}\right)
\nonumber \\ [.3cm] & &
\delta\left( 2\left[ n_{21}+n_{22}+n_{23}\right] +
\ell_{21}+\ell_{22}+\ell_{23}\; ,\; 2n_{2}+\ell_{2}\right)
\nonumber \\ [.3cm] & &
\delta\left( 2\left[ n_{31}+n_{32}+n_{33}\right] +
\ell_{31}+\ell_{32}+\ell_{33}\; ,\; 2n_{r}+\ell_{r}\right)
\nonumber \\ [.3cm] & &
\left(\begin{array}{ccc|c}
\ell_{11} & \ell_{21} & \ell_{31} & \ell\\ [.1cm]
m_{11} & m_{21} & m_{31} & m_{\ell}\end{array}\right)
\left(\begin{array}{ccc|c}
\ell_{12} & \ell_{22} & \ell_{32} & 1\\ [.1cm]
m_{12} & m_{22} & m_{32} & m\end{array}\right)
\left(\begin{array}{ccc|c}
\ell_{13} & \ell_{23} & \ell_{33} & 0\\ [.1cm]
m_{13} & m_{23} & m_{33} & 0\end{array}\right)
\nonumber \\ [.3cm] & &
\left(\begin{array}{ccc|c}
\ell_{11} & \ell_{12} & \ell_{13} & \ell_{1}\\ [.1cm]
m_{11} & m_{12} & m_{13} & m_{1}\end{array}\right)
\left(\begin{array}{ccc|c}
\ell_{21} & \ell_{22} & \ell_{23} & \ell_{2}\\ [.1cm]
m_{21} & m_{22} & m_{23} & m_{2}\end{array}\right)
\left(\begin{array}{ccc|c}
\ell_{31} & \ell_{32} & \ell_{33} & \ell_{r}\\ [.1cm]
m_{31} & m_{32} & m_{33} & m_{r}\end{array}\right)\; ,
\label{rearformula}
\end{eqnarray}

\noindent
where the angular-momenta recoupling coefficients are defined by

\begin{eqnarray}
 & & \;\;\;\;\;
\left(\begin{array}{ccc|c}
\ell_{1} & \ell_{2} & \ell_{3} & \ell\\ [.1cm]
m_{1} & m_{2} & m_{3} & m\end{array}\right)
\; =
\nonumber \\ [.3cm] & = &
\sum_{L,M}\;
\Clebsch{\ell_{1}}{\ell_{2}}{L}{m_{1}}{m_{2}}{M}
\Clebsch{L}{\ell_{3}}{\ell}{M}{m_{3}}{m}
\Clebsch{\ell_{1}}{\ell_{2}}{L}{0}{0}{0}
\Clebsch{L}{\ell_{3}}{\ell}{0}{0}{0}\; ,
\end{eqnarray}

\noindent
and where the \bm{\alpha}-matrices, for the case of equal constituent flavour
masses, are given by

\begin{equation}
\bm{\alpha}_{ABC}\; =\;
\left(\begin{array}{ccc}
\fsml{1}{2} & \fsml{1}{2} & -\sqrt{\fsml{1}{2}}\\ [.1cm]
\fsml{1}{2} & \fsml{1}{2} & \sqrt{\fsml{1}{2}}\\ [.1cm]
-\sqrt{\fsml{1}{2}} & \sqrt{\fsml{1}{2}} & 0\end{array}\right)
\;\;\;\xrm{and}\;\;\;
\bm{\alpha}_{BAC}\; =\;
\left(\begin{array}{ccc}
\fsml{1}{2} & \fsml{1}{2} & \sqrt{\fsml{1}{2}}\\ [.1cm]
\fsml{1}{2} & \fsml{1}{2} & -\sqrt{\fsml{1}{2}}\\ [.1cm]
\sqrt{\fsml{1}{2}} & -\sqrt{\fsml{1}{2}} & 0\end{array}\right)\; .
\label{alphamat}
\end{equation}

\noindent
The allowed values for the quantum numbers of the internal lines of diagram
(\ref{reardiagram}), given by $n_{ij}$ and $\ell_{ij}$
in formula (\ref{rearformula}),
are non-negative integers and hence, because of partial quantum number
conservation at each vertex of the diagram, which is moreover
expressed by the Kronecker delta's in formula (\ref{rearformula}), we find

\begin{equation}
n_{12}\; =\; 
n_{22}\; =\; 
n_{32}\; =\; 
n_{13}\; =\; 
n_{23}\; =\; 
n_{33}\; =\;
\ell_{13}\; =\; 
\ell_{23}\; =\; 
\ell_{33}\; =\; 0\;\;\; ,
\end{equation}

\noindent
which, also substituting the \bm{\alpha}-matrices (\ref{alphamat}),
simplifies the expression for the rearrangement coefficients to

\begin{eqnarray}
 & & \;\;\;\;\;
\left(\begin{array}{ccccccc}
n & \ell & m_{\ell} &  & n_{1} & \ell_{1} & m_{1}\\ [.1cm]
0 & 1 & m &  & n_{2} & \ell_{2} & m_{2}\\ [.1cm]
0 & 0 & 0 &  & n_{r} & \ell_{r} & m_{r}
\end{array}\right)_\x{
\left\{\begin{array}{c}\bm{\alpha}_{ABC}\\\bm{\alpha}_{BAC}\end{array}
\right\} } =
(-1)^\x{n+n_{1}+n_{2}+n_{r}}\left(\fnd{\pi}{4}\right)^{3}
\sqrt{\left( n!\; n_{1}!\; n_{2}!\; n_{r}!\right)}
\nonumber \\ [.3cm] & &
\sqrt{\left(\fnd{
\Gamma\left( 2n+\ell+\fsml{3}{2}\right)
\Gamma\left(\fsml{5}{2}\right)
\Gamma\left(\fsml{3}{2}\right)
\Gamma\left( 2n_{1}+\ell_{1}+\fsml{3}{2}\right)
\Gamma\left( 2n_{2}+\ell_{2}+\fsml{3}{2}\right)
\Gamma\left( 2n_{r}+\ell_{r}+\fsml{3}{2}\right)}
{\left( 2\ell +1\right)
\left( 3\right)
\left( 2\ell_{1}+1\right)
\left( 2\ell_{2}+1\right)
\left( 2\ell_{r}+1\right)}\right)}
\nonumber \\ [.3cm] & &
\left(\fsml{1}{2}\right)^\x{2n+\ell +1-n_{r}-\fsml{1}{2}\ell_{r}}\;\;
\sum_\x{\left\{ n_{ij},\ell_{ij},m_{ij}\right\}}\;\;
\left\{\begin{array}{c} (-1)^\x{\ell_{31}}\\[.5cm]
(-1)^\x{\ell_{32}}\end{array}\right\}
\nonumber \\ [.3cm] & &
\fnd{1}{n_{11}!\;n_{21}!\;n_{31}!}\;
\fnd{
\left( 2\ell_{11}+1\right)
\left( 2\ell_{12}+1\right)
}
{
\Gamma\left( n_{11}+\ell_{11}+\fsml{3}{2}\right)
\Gamma\left(\ell_{12}+\fsml{3}{2}\right)
\Gamma\left(\fsml{3}{2}\right)
}
\nonumber \\ [.3cm] & &
\fnd{
\left( 2\ell_{21}+1\right)
\left( 2\ell_{22}+1\right)
}
{
\Gamma\left( n_{21}+\ell_{21}+\fsml{3}{2}\right)
\Gamma\left(\ell_{22}+\fsml{3}{2}\right)
\Gamma\left(\fsml{3}{2}\right)
}\;
\fnd{
\left( 2\ell_{31}+1\right)
\left( 2\ell_{32}+1\right)
}
{
\Gamma\left( n_{31}+\ell_{31}+\fsml{3}{2}\right)
\Gamma\left(\ell_{32}+\fsml{3}{2}\right)
\Gamma\left(\fsml{3}{2}\right)
}
\nonumber \\ [.3cm] & &
\delta\left( 2\left[ n_{11}+n_{21}+n_{31}\right] +
\ell_{11}+\ell_{21}+\ell_{31}\; ,\; 2n+\ell\right)\;
\delta\left( \ell_{12}+\ell_{22}+\ell_{32}\; ,\; 1\right)
\nonumber \\ [.3cm] & &
\delta\left( 2n_{11} +\ell_{11}+\ell_{12}\; ,\; 2n_{1}+\ell_{1}\right)\;
\delta\left( 2n_{21} +\ell_{21}+\ell_{22}\; ,\; 2n_{2}+\ell_{2}\right)
\nonumber \\ [.3cm] & &
\delta\left( 2n_{31} +
\ell_{31}+\ell_{32}\; ,\; 2n_{r}+\ell_{r}\right)
\nonumber \\ [.3cm] & &
\left(\begin{array}{ccc|c}
\ell_{11} & \ell_{21} & \ell_{31} & \ell\\ [.1cm]
m_{11} & m_{21} & m_{31} & m_{\ell}\end{array}\right)
\Clebsch{\ell_{11}}{\ell_{12}}{\ell_{1}}{m_{11}}{m_{12}}{m_{1}}
\Clebsch{\ell_{11}}{\ell_{12}}{\ell_{1}}{0}{0}{0}
\nonumber \\ [.3cm] & &
\Clebsch{\ell_{21}}{\ell_{22}}{\ell_{2}}{m_{21}}{m_{22}}{m_{2}}
\Clebsch{\ell_{21}}{\ell_{22}}{\ell_{2}}{0}{0}{0}
\Clebsch{\ell_{31}}{\ell_{32}}{\ell_{r}}{m_{31}}{m_{32}}{m_{r}}
\Clebsch{\ell_{31}}{\ell_{32}}{\ell_{r}}{0}{0}{0}
\; .
\label{rrrngmnt}
\end{eqnarray}

\noindent
Notice that the Kronecker deltas in formula (\ref{rrrngmnt}) amount to

\begin{equation}
2n+\ell +1\; =\; 
2\left( n_{1}+n_{2}+n_{r}\right) +\ell_{1}+\ell_{2}+\ell_{r}\;\;\; ,
\label{finstat}
\end{equation}

\noindent
which is precisely the important relation that limits the number of possible
decay channels.
\vspace{0.5cm}

Moreover, for an initial pseudoscalar or vector meson out of the lowest-lying
flavour nonets, one has in formula (\ref{decint})
for the $q\bar{q}$ quantum numbers $n$ and $\ell$ that

\begin{equation}
n\; =\;\ell\; =\; 0\;\;\; .
\end{equation}

\noindent
Consequently, through the use of the Kronecker deltas in Eq.~(\ref{rrrngmnt}),
we find for the quantum numbers of the internal lines of diagram
(\ref{reardiagram}) that

\begin{equation}
n_{11}\; =\; n_{21}\; =\; n_{31}\; =\;
\ell_{11}\; =\; \ell_{21}\; =\; \ell_{31}\; =\; 0
\;\;\; ,
\end{equation}

\noindent
and moreover

\begin{equation}
n_{1}\; =\; n_{2}\; =\; n_{r}\; =\; 0
\;\;\;\xrm{and}\;\;\;
\ell_{1}\; +\; \ell_{2}\; +\; \ell_{r}\; =\; 1
\;\;\; .
\label{psvecrel}
\end{equation}

\noindent
Relations (\ref{psvecrel}) can be checked against the first three colums of
Tables (\ref{pseudoscalar}) and (\ref{vector}), where
for all possible channels
the radial excitations $n_{1}$, $n_{2}$, or $n_{r}$ vanish,
and, moreover, the sums of $\ell_{1}$, $\ell_{2}$, and $\ell_{r}$ equal 1.
This is a consequence of formula (\ref{finstat}) and drastically limits the
number of possible quantum numbers and hence decay channels.

For the relevant rearrangement coefficients we find, using formula
(\ref{rrrngmnt}), in this case

\begin{equation}
\left(\begin{array}{ccccccc}
0 & 0 & 0 &  & 0 & \ell_{1} & m_{1}\\ [.1cm]
0 & 1 & m &  & 0 & \ell_{2} & m_{2}\\ [.1cm]
0 & 0 & 0 &  & 0 & \ell_{r} & m_{r}
\end{array}\right)_\x{
\left\{\begin{array}{c}\bm{\alpha}_{ABC}\\\bm{\alpha}_{BAC}\end{array}
\right\} }\; =\;
\left(\fsml{1}{2}\right)^\x{1-\fsml{1}{2}\ell_{r}}\;\;
\left\{\begin{array}{c} +1\\[.5cm] (-1)^\x{\ell_{r}}\end{array}\right\}\;
\delta\left( \ell_{1}+\ell_{2}+\ell_{r}\; ,\; 1\right)\; .
\label{rearps+v}
\end{equation}

\noindent
For the decay of a meson out of the lowest-lying scalar nonet, one has

\begin{equation}
n\; =\; 0\;\;\;\xrm{and}\;\;\;\ell\; =\; s\; =\; 1\;\;\; ,
\end{equation}

\noindent
and hence, by the use of formula (\ref{rrrngmnt}),
one obtains for the relevant rearrangement coefficients in this case

\begin{eqnarray}
 & & \;\;\;\;\;
\left(\begin{array}{ccccccc}
0 & 1 & m_{\ell} &  & n_{1} & \ell_{1} & m_{1}\\ [.1cm]
0 & 1 & m &  & n_{2} & \ell_{2} & m_{2}\\ [.1cm]
0 & 0 & 0 &  & n_{r} & \ell_{r} & m_{r}
\end{array}\right)_\x{
\left\{\begin{array}{c}\bm{\alpha}_{ABC}\\\bm{\alpha}_{BAC}\end{array}
\right\} }\; =
\nonumber \\ [.3cm] & = &
(-1)^\x{n_{1}+n_{2}+n_{r}}\; 8\;
\sqrt{\left( n_{1}!\; n_{2}!\; n_{r}!\right)}
\nonumber \\ [.3cm] & &
\sqrt{\left(\fnd{
\Gamma\left( 2n_{1}+\ell_{1}+\fsml{3}{2}\right)
\Gamma\left( 2n_{2}+\ell_{2}+\fsml{3}{2}\right)
\Gamma\left( 2n_{r}+\ell_{r}+\fsml{3}{2}\right)}
{2\pi^{3/2}\;
\left( 2\ell_{1}+1\right)
\left( 2\ell_{2}+1\right)
\left( 2\ell_{r}+1\right)}\right)}
\nonumber \\ [.3cm] & &
\left(\fsml{1}{2}\right)^\x{2n_{1}+2n_{2}+n_{r}+\ell_{1}+\ell_{2}
+\fsml{1}{2}\ell_{r}}\;\;
\sum_\x{\left\{ n_{ij},\ell_{ij},m_{ij}\right\}}\;\;
\left\{\begin{array}{c} (-1)^\x{\ell_{31}}\\[.5cm]
(-1)^\x{\ell_{32}}\end{array}\right\}
\nonumber \\ [.3cm] & &
\delta\left(\ell_{11}+\ell_{21}+\ell_{31}\; ,\; 1\right)\;
\delta\left(\ell_{12}+\ell_{22}+\ell_{32}\; ,\; 1\right)
\nonumber \\ [.3cm] & &
\delta\left(\ell_{11}+\ell_{12}\; ,\; 2n_{1}+\ell_{1}\right)\;
\delta\left(\ell_{21}+\ell_{22}\; ,\; 2n_{2}+\ell_{2}\right)
\delta\left(\ell_{31}+\ell_{32}\; ,\; 2n_{r}+\ell_{r}\right)
\nonumber \\ [.3cm] & &
\Clebsch{\ell_{11}}{\ell_{12}}{\ell_{1}}{m_{11}}{m_{12}}{m_{1}}
\Clebsch{\ell_{11}}{\ell_{12}}{\ell_{1}}{0}{0}{0}
\nonumber \\ [.3cm] & &
\Clebsch{\ell_{21}}{\ell_{22}}{\ell_{2}}{m_{21}}{m_{22}}{m_{2}}
\Clebsch{\ell_{21}}{\ell_{22}}{\ell_{2}}{0}{0}{0}
\Clebsch{\ell_{31}}{\ell_{32}}{\ell_{r}}{m_{31}}{m_{32}}{m_{r}}
\Clebsch{\ell_{31}}{\ell_{32}}{\ell_{r}}{0}{0}{0}
\; .
\label{rearscalar}
\end{eqnarray}

\section{Mixing for scalar mesons}
\label{scalarmixing}

Since we assume $^{3}P_{0}$ quantum numbers for the creation of a
$q\bar{q}$ pair out of the vacuum, the mixing of the isoscalar flavour-nonet
members for lowest-lying scalar meson decay is not completely trivial.
So we will outline here some of the necessary ingredients.

In order to simplify the discussion, let us separate the normalisation
factor and the summations in  formula (\ref{decABC}).
Moreover, the two Kronecker deltas under the square root in formula
(\ref{decABC}) do not vanish for the lowest-lying scalar mesons.
Therefore, let us denote

\begin{eqnarray}
 & & \;\;\;\;\;\left.\begin{array}{l}
\bracket{0,0,j_{1},\ell_{1},s_{1},n_{1},
j_{2},\ell_{2},s_{2},n_{2},\ell_{r},s_{r},n_{r}}
{0,0,1,1,0,\bm{\alpha}_{ABC}}\\ [.5cm]
\bracket{0,0,j_{1},\ell_{1},s_{1},n_{1},
j_{2},\ell_{2},s_{2},n_{2},\ell_{r},s_{r},n_{r}}
{0,0,1,1,0,\bm{\alpha}_{BAC}}\end{array}\right\}\; =
\label{decsimp} \\ [.5cm] & &
=\;\fnd{1}{\sqrt{1+
\bracket{C}{SU(3)_\xrm{\scriptsize flavour}\xrm{-singlet}}}}\;\;
\times\;\left\{\begin{array}{l} \valexp{ABC} \\ [.3cm] \valexp{BAC}
\end{array}\right.
\;\;\;\; ,
\nonumber
\end{eqnarray}

\noindent
where $\valexp{ABC}$ stands for the upper summation in Eq.~(\ref{decABC})
and $\valexp{BAC}$ for the lower.

As one notices from formulae (\ref{decABC}), (\ref{reardiagram}),
(\ref{rearformula}) or just only from formula (\ref{rearscalar}), the
transition coefficients $\valexp{ABC}$ and $\valexp{BAC}$ do not ,
for full $SU(3)$ flavour symmetry, i.e., for equal up, down and strange
quark masses,
depend on the flavour contents of the three mesons $A$, $B$, and $C$
involved in the transition process (\ref{MMdecay}), but just on the orbital
and intrinsic spin quantum numbers of the system, which circumstance is also
expressed by the notation of formula (\ref{spatial}).
In fact, for the case of equal quark masses, those transition coefficients
are equal for each different set of spatial quantum numbers, up to a sign.
This sign is positive for all possible couplings in the case of
the lowest-lying scalar mesons.
Consequently, since for Table \ref{tabspsps} only the transitions between
scalar mesons and pairs of pseudoscalar mesons are relevant, and for mixing
in general only the restriction to a specific set of spatial quantum numbers
has to be considered, we may put here

\begin{equation}
\valexp{ABC}\; =\;\valexp{BAC}
\;\;\; .
\label{ABCequal}
\end{equation}

For the transitions of the flavour-octet members to pairs of mesons, one has,
according to formula (\ref{normtab}), a unity normalisation factor.
Let us study then the matrix elements for a representant, $u\bar{d}$,
of the isotriplets, which is denoted by $t$ in Table \ref{scalar},
coupled to an isoscalar, $\phi$, and an isotriplet, for which we also take
as a representant the $u\bar{d}$ state, and which, moreover, is also denoted
by $t$ in Table \ref{scalar}, i.e.,

\begin{equation}
\bracket{\left( u\bar{d}\right)\phi}{u\bar{d}}
\;\;\; .
\label{isotriplet}
\end{equation}

\noindent
If $\phi$ represents the flavour-octet-member isoscalar $\phi_{8}$,
we find

\begin{displaymath}
\bracket{\left( u\bar{d}\right)\phi_{8}}{u\bar{d}}\; =\;
\bracket{\left( u\bar{d}\right)\sqrt{\fsml{1}{6}}
\left( u\bar{u}+d\bar{d}-2s\bar{s}\right)}{u\bar{d}}
\;\;\; .
\end{displaymath}

\noindent
Obviously, the matrix element for the $s\bar{s}$ contribution to $\phi_{8}$
vanishes, hence

\begin{displaymath}
\bracket{\left( u\bar{d}\right)\phi_{8}}{u\bar{d}}\; =\;
\sqrt{\fsml{1}{6}}\left\{
\bracket{\left( u\bar{d}\right)\left( u\bar{u}\right)}{u\bar{d}}\; +\;
\bracket{\left( u\bar{d}\right)\left( d\bar{d}\right)}{u\bar{d}}
\right\}
\;\;\; ,
\end{displaymath}

\noindent
which, also using Eq.~(\ref{ABCequal}), gives

\begin{equation}
\bracket{\left( u\bar{d}\right)\phi_{8}}{u\bar{d}}\; =\;
\sqrt{\fsml{1}{6}}\left\{\valexp{BAC}+\valexp{ABC}\right\}
\; =\;\sqrt{\fsml{2}{3}}\valexp{ABC}
\;\;\; .
\label{isot8}
\end{equation}

Following a similar reasoning when $\phi$ in Eq.~(\ref{isotriplet})
represents the flavour-singlet isoscalar $\phi_{1}$, we find for its matrix
elements the result 

\begin{equation}
\bracket{\left( u\bar{d}\right)\phi_{1}}{u\bar{d}}\; =\;
\bracket{\left( u\bar{d}\right)
\sqrt{\fsml{1}{3}}\left( u\bar{u}+d\bar{d}+s\bar{s}\right)}
{u\bar{d}}\; =\;
\sqrt{\fsml{4}{3}}\valexp{ABC}
\;\;\; .
\label{isot1}
\end{equation}

\noindent
In Table \ref{scalar}, for the quadratic matrix elements under $t8$ and
$t1$ in the sector {\it isotriplets}, one may verify the factor 2
that follows from formulae (\ref{isot8}) and (\ref{isot1}),

For the ideally-mixed isoscalars, $\phi_{n}$ (non-strange) and
$\phi_{s}$ (strange), defined by

\begin{equation}
\phi_{n}\; =\;
\sqrt{\fsml{1}{2}}\left( u\bar{u}+d\bar{d}\right)\; =\;
\sqrt{\fsml{2}{3}}\phi_{1}+
\sqrt{\fsml{1}{3}}\phi_{8}
\;\;\;\xrm{and}\;\;\;
\phi_{s}\; =\; s\bar{s}\; =\;
\sqrt{\fsml{1}{3}}\phi_{1}-
\sqrt{\fsml{2}{3}}\phi_{8}
\;\;\; ,
\label{idealmix}
\end{equation}

\noindent
one finds the matrix elements

\begin{eqnarray}
\bracket{\left( u\bar{d}\right)\phi_{n}}{u\bar{d}} & = &
\sqrt{2}\valexp{ABC}
\; =\;\sqrt{3}\bracket{\left( u\bar{d}\right)\phi_{8}}{u\bar{d}}
\; =\;\sqrt{\fsml{3}{2}}\bracket{\left( u\bar{d}\right)\phi_{1}}{u\bar{d}}
\;\;\;\xrm{and}
\nonumber \\ [.3cm]
\bracket{\left( u\bar{d}\right)\phi_{s}}{u\bar{d}}
 & = & 0\;\;\; .
\label{isotns}
\end{eqnarray}

\noindent
Besides the multiplicative factor of 24 which is discussed in Section
\ref{comparison}, formula (\ref{isotns}) establishes the relation between
the values given in the first line of Table \ref{scalar} and the values
given in Table \ref{tabspsps} for the following matrix elements:

\begin{equation}
\bracket{\pi\eta_{n}}{a_{0}}\; =\;
\sqrt{24}\sqrt{3}\bracket{t8}{t}\; =\;
\sqrt{\fsml{2}{3}}
\;\;\;\xrm{and}\;\;\;
\bracket{\pi\eta_{s}}{a_{0}}\; =\; 0
\;\;\; .
\label{ppea0}
\end{equation}

For the coupling of an isodoublet lowest-lying scalar meson to 
the isodoublet-isoscalar pair, we may also select representants.
Let us consider the matrix element

\begin{displaymath}
\bracket{\left( u\bar{s}\right)\phi}{u\bar{s}}
\;\;\; .
\end{displaymath}

\noindent
In this case, the matrix element for the $d\bar{d}$ contribution vanishes.
Consequently, for the isoscalar flavour singlets and octets, we end up with

\begin{equation}
\bracket{\left( u\bar{s}\right)\phi_{1}}{u\bar{s}}\; =\;
\sqrt{\fsml{4}{3}}\valexp{ABC}
\;\;\;\xrm{and}\;\;\;
\bracket{\left( u\bar{s}\right)\phi_{8}}{u\bar{s}}\; =\;
-\sqrt{\fsml{1}{6}}\valexp{ABC}
\;\;\; ,
\label{isod18}
\end{equation}

\noindent
which explains the factor 8 in Table \ref{scalar} for the
quadratic matrix elements under $d8$ and $d1$, in the sector under
{\it isodoublets}.

From Eq.~(\ref{isod18}), we obtain for the ideally mixed combinations
(\ref{idealmix}) the relations

\begin{eqnarray*}
\bracket{\left( u\bar{s}\right)\phi_{n}}{u\bar{s}} & = &
\sqrt{\fsml{1}{2}}\valexp{BAC}\; =\;
\sqrt{\fsml{3}{8}}\bracket{\left( u\bar{s}\right)\phi_{1}}{u\bar{s}}
\;\;\;\xrm{and}
 \\ [.3cm]
\bracket{\left( u\bar{s}\right)\phi_{s}}{u\bar{s}} & = &
\valexp{ABC}\; =\;
\sqrt{\fsml{3}{4}}\bracket{\left( u\bar{s}\right)\phi_{1}}{u\bar{s}}
\;\;\; ,
\end{eqnarray*}

\noindent
which, by the use of the first line of Table \ref{scalar} and when, moreover,
multiplied by the factor $\sqrt{24}$, gives the matrix elements of
$\kappa\rightarrow K\eta_{n}$ and $\kappa\rightarrow K\eta_{s}$, i.e.,

\begin{displaymath}
\bracket{K\eta_{n}}{\kappa}\; =\;
\sqrt{24}\sqrt{\fsml{3}{8}}
\bracket{d1}{d}\; =\;
\sqrt{\fsml{1}{6}}
\;\;\;\xrm{and}\;\;\;
\bracket{K\eta_{s}}{\kappa}\; =\;
\sqrt{24}\sqrt{\fsml{3}{4}}
\bracket{d1}{d}\; =\;
\sqrt{\fsml{1}{3}}
\;\;\; ,
\end{displaymath}

\noindent
the quadratic sum ($=\displaystyle\fsml{1}{2}$) of which is found for model
B86 in Table \ref{tabspsps}.

Now, according to formula (\ref{normtab}), for the couplings of lowest-lying
scalar isoscalars to meson pairs, the normalisation may be not unity.
Let us begin with the coupling to a pair of isotriplets, e.g.

\begin{displaymath}
\bracket{\left( u\bar{d}\right)\left( d\bar{u}\right)}{\phi}
\;\;\; .
\end{displaymath}

\noindent
When $\phi$ represents a flavour-singlet isoscalar, then, also using
formulae (\ref{normtab}) and (\ref{ABCequal}), one finds

\begin{displaymath}
\bracket{\left( u\bar{d}\right)\left( d\bar{u}\right)}{\phi_{1}}=
\bracket{\left( u\bar{d}\right)\left( d\bar{u}\right)}
{\sqrt{\fsml{1}{3}}\left( u\bar{u}+d\bar{d}+s\bar{s}\right)}=
\sqrt{\fsml{1}{3}}
\left\{\fnd{\valexp{ABC}}{\sqrt{2}}+\fnd{\valexp{BAC}}{\sqrt{2}}\right\}
=\sqrt{\fsml{2}{3}}\valexp{ABC}
\;\;\; ,
\end{displaymath}

\noindent
whereas, when $\phi$ represents a flavour-octet isoscalar, it follows that

\begin{displaymath}
\bracket{\left( u\bar{d}\right)\left( d\bar{u}\right)}{\phi_{8}}\; =\;
\sqrt{\fsml{1}{6}}
\left\{\valexp{ABC}+\valexp{BAC}\right\}\; =\;
\sqrt{\fsml{2}{3}}\valexp{ABC}
\;\;\; .
\end{displaymath}

\noindent
Indeed, in Table \ref{scalar} the quadratic matrix elements under $tt$
in the sector for the flavour-octet isoscalars are the same as in the sector
for the flavour-singlet isoscalars.

For the ideally mixed combination (\ref{idealmix}), also applying
formulae (\ref{normtab}) and (\ref{ABCequal}), one obtains

\begin{equation}
\bracket{\left( u\bar{d}\right)\left( d\bar{u}\right)}{\phi_{n}}\; =\;
\sqrt{\fsml{1}{2}}
\left\{\fnd{\valexp{ABC}}{\sqrt{\fsml{5}{3}}}+
\fnd{\valexp{BAC}}{\sqrt{\fsml{5}{3}}}\right\}
\; =\;
\sqrt{\fsml{6}{5}}\valexp{ABC}
\; =\;
\sqrt{\fsml{9}{5}}
\bracket{\left( u\bar{d}\right)\left( d\bar{u}\right)}{\phi_{1}}
\;\;\; .
\label{isoscttn}
\end{equation}

\noindent
When, moreover, multiplied by the factor $\sqrt{24}$, formula
(\ref{isoscttn}) establishes the relation between the first line of Table
\ref{scalar} and the matrix element of $\eta_{n}\rightarrow\pi\pi$,
according to

\begin{displaymath}
\bracket{\pi\pi}{\eta_{n}}\; =\;
\sqrt{24}\sqrt{\fsml{9}{5}}
\bracket{tt}{\phi_{1}}\; =\;
\sqrt{\fsml{3}{5}}
\;\;\; ,
\end{displaymath}

\noindent
as is found for model B86 in Table \ref{tabspsps}.

Next, let us also study the coupling of isoscalars to a pair of isodoublets,
e.g.

\begin{displaymath}
\bracket{\left( u\bar{s}\right)\left( s\bar{u}\right)}{\phi}
\;\;\; .
\end{displaymath}

When $\phi$ represents a flavour-singlet isoscalar, then, again through the use
of formulae (\ref{normtab}) and (\ref{ABCequal}), we have

\begin{displaymath}
\bracket{\left( u\bar{s}\right)\left( s\bar{u}\right)}{\phi_{1}}\; =\;
\sqrt{\fsml{1}{3}}
\left\{\fnd{\valexp{ABC}}{\sqrt{2}}+\fnd{\valexp{BAC}}{\sqrt{2}}\right\}
\; =\;
\sqrt{\fsml{2}{3}}\valexp{ABC}
\;\;\; .
\end{displaymath}

\noindent
Similarly, for a flavour-octet isoscalar we find

\begin{displaymath}
\bracket{\left( u\bar{s}\right)\left( s\bar{u}\right)}{\phi_{8}}\; =\;
\sqrt{\fsml{1}{6}}
\left\{\valexp{ABC}-2\valexp{BAC}\right\}
\; =\; -\sqrt{\fsml{1}{6}}\valexp{ABC}
\;\;\; ,
\end{displaymath}

\noindent
which result explains the factor $\fsml{1}{4}$ in Table \ref{scalar}
between the quadratic matrix elements under $dd$ in the sectors for the
flavour-octet isoscalars and for the flavour-singlet isoscalars.

For the ideally mixed isoscalars defined in formula (\ref{idealmix}),
using once again formula (\ref{normtab}), we obtain 

\begin{eqnarray*}
\bracket{\left( u\bar{s}\right)\left( s\bar{u}\right)}{\phi_{n}} & = &
\sqrt{\fsml{1}{2}}\fnd{\valexp{ABC}}{\sqrt{\fsml{5}{3}}}
\; =\;\sqrt{\fsml{3}{10}}\valexp{ABC}\; =\;
\sqrt{\fsml{9}{20}}
\bracket{\left( u\bar{s}\right)\left( s\bar{u}\right)}{\phi_{1}}
\;\;\;\xrm{and} \\ [.3cm]
\bracket{\left( u\bar{s}\right)\left( s\bar{u}\right)}{\phi_{s}} & = &
\sqrt{\fsml{1}{2}}\fnd{\valexp{BAC}}{\sqrt{\fsml{4}{3}}}
\; =\;\sqrt{\fsml{3}{4}}\valexp{ABC}\; =\;
\sqrt{\fsml{9}{8}}
\bracket{\left( u\bar{s}\right)\left( s\bar{u}\right)}{\phi_{1}}
\;\;\; ,
\end{eqnarray*}

\noindent
which nicely explains the values for the matrix elements of
$\left(\eta_{n}/\eta_{s}\right)\rightarrow K\bar{K}$, i.e.,

\begin{displaymath}
\bracket{K\bar{K}}{\eta_{n}}\; =\;
\sqrt{24}\sqrt{\fsml{9}{20}}
\bracket{dd}{\phi_{1}}\; =\;
\sqrt{\fsml{1}{5}}
\;\;\;\xrm{and}\;\;\;
\bracket{K\bar{K}}{\eta_{s}}\; =\;
\sqrt{24}\sqrt{\fsml{9}{8}}
\bracket{dd}{\phi_{1}}\; =\;
\sqrt{\fsml{1}{2}}
\;\;\; ,
\end{displaymath}

\noindent
given for model B86 in Table \ref{tabspsps}.

Similar straightforward calculations lead to the matrix elements
$\bracket{\eta_{n}\eta_{n}}{\eta_{n}}$ and
$\bracket{\eta_{s}\eta_{s}}{\eta_{s}}$.

\clearpage

\begin{table}[h]
\begin{center}
\begin{tabular}{||c|c||}
\hline\hline & \\
symbol & multiplet \\
\hline\hline & \\
$t$ & isotriplets\\ [.3cm]
$d$ & isodoublets\\ [.3cm]
$8$ & isoscalar $SU_{3}$-octet members\\ [.3cm]
$1$ & $SU_{3}$-singlets\\ \hline\hline
\end{tabular}
\end{center}
\caption[]{Particle identification used in this paper.}
\label{partid}
\end{table}
\clearpage

\begin{table}[h]
\begin{center}
\scriptsize
\begin{tabular}{||cc|c||cccc|c||ccc|c||cccc|c||cccc|c||}
\cline{4-22} \multicolumn{3}{c||}{} & \multicolumn{19}{c||}{} \\
\multicolumn{3}{c||}{} &
\multicolumn{19}{c||}{flavour channels and totals for $M(J\ell sn\; =\;0000)$}\\
\cline{4-22} \multicolumn{3}{c||}{} &
\multicolumn{14}{c||}{} & \multicolumn{5}{c||}{} \\
\multicolumn{3}{c||}{} &
\multicolumn{14}{c||}{$SU_{3}$-octet members} &
\multicolumn{5}{c||}{$SU_{3}$-singlets} \\
\hline \multicolumn{3}{||c||}{} &
\multicolumn{14}{c||}{} & \multicolumn{5}{c||}{} \\
\multicolumn{3}{||c||}{spatial q-numbers} & \multicolumn{5}{c||}{isotriplets} &
\multicolumn{4}{c||}{isodoublets} &
\multicolumn{5}{c||}{isoscalars} & \multicolumn{5}{c||}{} \\ [.3cm]
$M_{1}$ & $M_{2}$ & rel. & \multicolumn{5}{c||}{($t$)} &
\multicolumn{4}{c||}{($d$)} & \multicolumn{5}{c||}{($8$)} &
\multicolumn{5}{c||}{($1$)}\\ [.3cm]
$\!\! j\ell sn\!\!$ & $\!\! j\ell sn\!\!$ & $\!\! \ell sn\!\!$ &
$\!\! tt\!\!$ & $\!\! dd\!\!$ & $\!\! t8\!\!$ &
$\!\! t1\!\!$ & $\!\! T\!\!$ & $\!\! td\!\!$ & $\!\! d8\!\!$ &
$\!\! d1\!\!$ & $\!\! T\!\!$ & $\!\! tt\!\!$ & $\!\! dd\!\!$ &
$\!\! 88\!\!$ & $\!\! 11\!\!$ & $\!\! T\!\!$ & $\!\! tt\!\!$ &
$\!\! dd\!\!$ & $\!\! 88\!\!$ & $\!\! 11\!\!$ & $\!\! T\!\!$ \\
\hline & & & & & & & & & & & & & & & & & & & & & \\ [.3cm]
$\!\! 0000\!\!$ & $\!\! 0110\!\!$ & $\!\! 000\!\!$ &
- &
$\!\!\frac{    1}{   24}\!\!$ &
$\!\!\frac{    1}{   36}\!\!$ &
$\!\!\frac{    1}{   18}\!\!$ &
$\!\!\frac{    1}{    8}\!\!$ &
$\!\!\frac{    1}{   16}\!\!$ &
$\!\!\frac{    1}{  144}\!\!$ &
$\!\!\frac{    1}{   18}\!\!$ &
$\!\!\frac{    1}{    8}\!\!$ &
$\!\!\frac{    1}{   24}\!\!$ &
$\!\!\frac{    1}{   72}\!\!$ &
$\!\!\frac{    1}{   72}\!\!$ &
$\!\!\frac{    1}{   18}\!\!$ &
$\!\!\frac{    1}{    8}\!\!$ &
$\!\!\frac{    1}{   12}\!\!$ &
$\!\!\frac{    1}{    9}\!\!$ &
$\!\!\frac{    1}{   36}\!\!$ &
$\!\!\frac{    1}{   36}\!\!$ &
$\!\!\frac{    1}{    4}\!\!$\\ [.3cm]
$\!\! 1010\!\!$ & $\!\! 1100\!\!$ & $\!\! 000\!\!$ &
- &
$\!\!\frac{    1}{   24}\!\!$ &
$\!\!\frac{    1}{   36}\!\!$ &
$\!\!\frac{    1}{   18}\!\!$ &
$\!\!\frac{    1}{    8}\!\!$ &
$\!\!\frac{    1}{   16}\!\!$ &
$\!\!\frac{    1}{  144}\!\!$ &
$\!\!\frac{    1}{   18}\!\!$ &
$\!\!\frac{    1}{    8}\!\!$ &
$\!\!\frac{    1}{   24}\!\!$ &
$\!\!\frac{    1}{   72}\!\!$ &
$\!\!\frac{    1}{   72}\!\!$ &
$\!\!\frac{    1}{   18}\!\!$ &
$\!\!\frac{    1}{    8}\!\!$ &
$\!\!\frac{    1}{   12}\!\!$ &
$\!\!\frac{    1}{    9}\!\!$ &
$\!\!\frac{    1}{   36}\!\!$ &
$\!\!\frac{    1}{   36}\!\!$ &
$\!\!\frac{    1}{    4}\!\!$\\ [.3cm]
$\!\! 1010\!\!$ & $\!\! 1110\!\!$ & $\!\! 000\!\!$ &
$\!\!\frac{    1}{    6}\!\!$ &
$\!\!\frac{    1}{   12}\!\!$ &
- &
- &
$\!\!\frac{    1}{    4}\!\!$ &
$\!\!\frac{    1}{    8}\!\!$ &
$\!\!\frac{    1}{    8}\!\!$ &
- &
$\!\!\frac{    1}{    4}\!\!$ &
- &
$\!\!\frac{    1}{    4}\!\!$ &
- &
- &
$\!\!\frac{    1}{    4}\!\!$ &
- &
- &
- &
- &
- \\ [.3cm]
$\!\! 0000\!\!$ & $\!\! 1010\!\!$ & $\!\! 110\!\!$ &
$\!\!\frac{    1}{    6}\!\!$ &
$\!\!\frac{    1}{   12}\!\!$ &
- &
- &
$\!\!\frac{    1}{    4}\!\!$ &
$\!\!\frac{    1}{    8}\!\!$ &
$\!\!\frac{    1}{    8}\!\!$ &
- &
$\!\!\frac{    1}{    4}\!\!$ &
- &
$\!\!\frac{    1}{    4}\!\!$ &
- &
- &
$\!\!\frac{    1}{    4}\!\!$ &
- &
- &
- &
- &
- \\ [.3cm]
$\!\! 1010\!\!$ & $\!\! 1010\!\!$ & $\!\! 110\!\!$ &
- &
$\!\!\frac{    1}{   12}\!\!$ &
$\!\!\frac{    1}{   18}\!\!$ &
$\!\!\frac{    1}{    9}\!\!$ &
$\!\!\frac{    1}{    4}\!\!$ &
$\!\!\frac{    1}{    8}\!\!$ &
$\!\!\frac{    1}{   72}\!\!$ &
$\!\!\frac{    1}{    9}\!\!$ &
$\!\!\frac{    1}{    4}\!\!$ &
$\!\!\frac{    1}{   12}\!\!$ &
$\!\!\frac{    1}{   36}\!\!$ &
$\!\!\frac{    1}{   36}\!\!$ &
$\!\!\frac{    1}{    9}\!\!$ &
$\!\!\frac{    1}{    4}\!\!$ &
$\!\!\frac{    1}{    6}\!\!$ &
$\!\!\frac{    2}{    9}\!\!$ &
$\!\!\frac{    1}{   18}\!\!$ &
$\!\!\frac{    1}{   18}\!\!$ &
$\!\!\frac{    1}{    2}\!\!$
 \\ & & & & & & & & & & & & & & & & & & & & & \\ \hline
\end{tabular}
\normalsize
\end{center}
\caption[]{Coupling constants for the decay processes of pseudoscalar
mesons into meson pairs. The interpretation of the content of the table is
explained in the text.}
\label{pseudoscalar}
\end{table}
\clearpage

\mbox{}
\begin{table}[h]
\begin{center}
\scriptsize
\begin{tabular}{||cc|c||cccc|c||ccc|c||cccc|c||cccc|c||}
\cline{4-22} \multicolumn{3}{c||}{} & \multicolumn{19}{c||}{} \\
\multicolumn{3}{c||}{} &
\multicolumn{19}{c||}{flavour channels and totals for $M(J\ell sn\; =\;1010)$}\\
\cline{4-22} \multicolumn{3}{c||}{} &
\multicolumn{14}{c||}{} & \multicolumn{5}{c||}{} \\
\multicolumn{3}{c||}{} &
\multicolumn{14}{c||}{$SU_{3}$-octet members} &
\multicolumn{5}{c||}{$SU_{3}$-singlets} \\
\hline \multicolumn{3}{||c||}{} &
\multicolumn{14}{c||}{} & \multicolumn{5}{c||}{} \\
\multicolumn{3}{||c||}{spatial q-numbers} & \multicolumn{5}{c||}{isotriplets} &
\multicolumn{4}{c||}{isodoublets} &
\multicolumn{5}{c||}{isoscalars} & \multicolumn{5}{c||}{} \\ [.3cm]
$M_{1}$ & $M_{2}$ & rel. & \multicolumn{5}{c||}{($t$)} &
\multicolumn{4}{c||}{($d$)} & \multicolumn{5}{c||}{($8$)} &
\multicolumn{5}{c||}{($1$)}\\ [.3cm]
$\!\! j\ell sn\!\!$ & $\!\! j\ell sn\!\!$ & $\!\! \ell sn\!\!$ &
$\!\! tt\!\!$ & $\!\! dd\!\!$ & $\!\! t8\!\!$ &
$\!\! t1\!\!$ & $\!\! T\!\!$ & $\!\! td\!\!$ & $\!\! d8\!\!$ &
$\!\! d1\!\!$ & $\!\! T\!\!$ & $\!\! tt\!\!$ & $\!\! dd\!\!$ &
$\!\! 88\!\!$ & $\!\! 11\!\!$ & $\!\! T\!\!$ & $\!\! tt\!\!$ &
$\!\! dd\!\!$ & $\!\! 88\!\!$ & $\!\! 11\!\!$ & $\!\! T\!\!$ \\
\hline & & & & & & & & & & & & & & & & & & & & & \\ [.3cm]
$\!\! 0000\!\!$ & $\!\! 1100\!\!$ & $\!\! 010\!\!$ &
- &
$\!\!\frac{    1}{   72}\!\!$ &
$\!\!\frac{    1}{  108}\!\!$ &
$\!\!\frac{    1}{   54}\!\!$ &
$\!\!\frac{    1}{   24}\!\!$ &
$\!\!\frac{    1}{   48}\!\!$ &
$\!\!\frac{    1}{  432}\!\!$ &
$\!\!\frac{    1}{   54}\!\!$ &
$\!\!\frac{    1}{   24}\!\!$ &
$\!\!\frac{    1}{   72}\!\!$ &
$\!\!\frac{    1}{  216}\!\!$ &
$\!\!\frac{    1}{  216}\!\!$ &
$\!\!\frac{    1}{   54}\!\!$ &
$\!\!\frac{    1}{   24}\!\!$ &
$\!\!\frac{    1}{   36}\!\!$ &
$\!\!\frac{    1}{   27}\!\!$ &
$\!\!\frac{    1}{  108}\!\!$ &
$\!\!\frac{    1}{  108}\!\!$ &
$\!\!\frac{    1}{   12}\!\!$\\ [.3cm]
$\!\! 0000\!\!$ & $\!\! 1110\!\!$ & $\!\! 010\!\!$ &
$\!\!\frac{    1}{   18}\!\!$ &
$\!\!\frac{    1}{   36}\!\!$ &
- &
- &
$\!\!\frac{    1}{   12}\!\!$ &
$\!\!\frac{    1}{   24}\!\!$ &
$\!\!\frac{    1}{   24}\!\!$ &
- &
$\!\!\frac{    1}{   12}\!\!$ &
- &
$\!\!\frac{    1}{   12}\!\!$ &
- &
- &
$\!\!\frac{    1}{   12}\!\!$ &
- &
- &
- &
- &
- \\ [.3cm]
$\!\! 1010\!\!$ & $\!\! 1100\!\!$ & $\!\! 010\!\!$ &
$\!\!\frac{    1}{   18}\!\!$ &
$\!\!\frac{    1}{   36}\!\!$ &
- &
- &
$\!\!\frac{    1}{   12}\!\!$ &
$\!\!\frac{    1}{   24}\!\!$ &
$\!\!\frac{    1}{   24}\!\!$ &
- &
$\!\!\frac{    1}{   12}\!\!$ &
- &
$\!\!\frac{    1}{   12}\!\!$ &
- &
- &
$\!\!\frac{    1}{   12}\!\!$ &
- &
- &
- &
- &
- \\ [.3cm]
$\!\! 1010\!\!$ & $\!\! 1110\!\!$ & $\!\! 010\!\!$ &
- &
$\!\!\frac{    1}{   18}\!\!$ &
$\!\!\frac{    1}{   27}\!\!$ &
$\!\!\frac{    2}{   27}\!\!$ &
$\!\!\frac{    1}{    6}\!\!$ &
$\!\!\frac{    1}{   12}\!\!$ &
$\!\!\frac{    1}{  108}\!\!$ &
$\!\!\frac{    2}{   27}\!\!$ &
$\!\!\frac{    1}{    6}\!\!$ &
$\!\!\frac{    1}{   18}\!\!$ &
$\!\!\frac{    1}{   54}\!\!$ &
$\!\!\frac{    1}{   54}\!\!$ &
$\!\!\frac{    2}{   27}\!\!$ &
$\!\!\frac{    1}{    6}\!\!$ &
$\!\!\frac{    1}{    9}\!\!$ &
$\!\!\frac{    4}{   27}\!\!$ &
$\!\!\frac{    1}{   27}\!\!$ &
$\!\!\frac{    1}{   27}\!\!$ &
$\!\!\frac{    1}{    3}\!\!$\\ [.3cm]
$\!\! 0110\!\!$ & $\!\! 1010\!\!$ & $\!\! 010\!\!$ &
- &
$\!\!\frac{    1}{   24}\!\!$ &
$\!\!\frac{    1}{   36}\!\!$ &
$\!\!\frac{    1}{   18}\!\!$ &
$\!\!\frac{    1}{    8}\!\!$ &
$\!\!\frac{    1}{   16}\!\!$ &
$\!\!\frac{    1}{  144}\!\!$ &
$\!\!\frac{    1}{   18}\!\!$ &
$\!\!\frac{    1}{    8}\!\!$ &
$\!\!\frac{    1}{   24}\!\!$ &
$\!\!\frac{    1}{   72}\!\!$ &
$\!\!\frac{    1}{   72}\!\!$ &
$\!\!\frac{    1}{   18}\!\!$ &
$\!\!\frac{    1}{    8}\!\!$ &
$\!\!\frac{    1}{   12}\!\!$ &
$\!\!\frac{    1}{    9}\!\!$ &
$\!\!\frac{    1}{   36}\!\!$ &
$\!\!\frac{    1}{   36}\!\!$ &
$\!\!\frac{    1}{    4}\!\!$\\ [.3cm]
$\!\! 0000\!\!$ & $\!\! 0000\!\!$ & $\!\! 100\!\!$ &
$\!\!\frac{    1}{   36}\!\!$ &
$\!\!\frac{    1}{   72}\!\!$ &
- &
- &
$\!\!\frac{    1}{   24}\!\!$ &
$\!\!\frac{    1}{   48}\!\!$ &
$\!\!\frac{    1}{   48}\!\!$ &
- &
$\!\!\frac{    1}{   24}\!\!$ &
- &
$\!\!\frac{    1}{   24}\!\!$ &
- &
- &
$\!\!\frac{    1}{   24}\!\!$ &
- &
- &
- &
- &
- \\ [.3cm]
$\!\! 1010\!\!$ & $\!\! 1010\!\!$ & $\!\! 100\!\!$ &
$\!\!\frac{    1}{  108}\!\!$ &
$\!\!\frac{    1}{  216}\!\!$ &
- &
- &
$\!\!\frac{    1}{   72}\!\!$ &
$\!\!\frac{    1}{  144}\!\!$ &
$\!\!\frac{    1}{  144}\!\!$ &
- &
$\!\!\frac{    1}{   72}\!\!$ &
- &
$\!\!\frac{    1}{   72}\!\!$ &
- &
- &
$\!\!\frac{    1}{   72}\!\!$ &
- &
- &
- &
- &
- \\ [.3cm]
$\!\! 0000\!\!$ & $\!\! 1010\!\!$ & $\!\! 110\!\!$ &
- &
$\!\!\frac{    1}{   18}\!\!$ &
$\!\!\frac{    1}{   27}\!\!$ &
$\!\!\frac{    2}{   27}\!\!$ &
$\!\!\frac{    1}{    6}\!\!$ &
$\!\!\frac{    1}{   12}\!\!$ &
$\!\!\frac{    1}{  108}\!\!$ &
$\!\!\frac{    2}{   27}\!\!$ &
$\!\!\frac{    1}{    6}\!\!$ &
$\!\!\frac{    1}{   18}\!\!$ &
$\!\!\frac{    1}{   54}\!\!$ &
$\!\!\frac{    1}{   54}\!\!$ &
$\!\!\frac{    2}{   27}\!\!$ &
$\!\!\frac{    1}{    6}\!\!$ &
$\!\!\frac{    1}{    9}\!\!$ &
$\!\!\frac{    4}{   27}\!\!$ &
$\!\!\frac{    1}{   27}\!\!$ &
$\!\!\frac{    1}{   27}\!\!$ &
$\!\!\frac{    1}{    3}\!\!$\\ [.3cm]
$\!\! 1010\!\!$ & $\!\! 1010\!\!$ & $\!\! 120\!\!$ &
$\!\!\frac{    5}{   27}\!\!$ &
$\!\!\frac{    5}{   54}\!\!$ &
- &
- &
$\!\!\frac{    5}{   18}\!\!$ &
$\!\!\frac{    5}{   36}\!\!$ &
$\!\!\frac{    5}{   36}\!\!$ &
- &
$\!\!\frac{    5}{   18}\!\!$ &
- &
$\!\!\frac{    5}{   18}\!\!$ &
- &
- &
$\!\!\frac{    5}{   18}\!\!$ &
- &
- &
- &
- &
- \\ & & & & & & & & & & & & & & & & & & & & & \\ \hline
\end{tabular}
\normalsize
\end{center}
\caption[]{Coupling constants for the decay processes of vector
mesons into meson pairs. The interpretation of the content of the table is
explained in the text.}
\label{vector}
\end{table}
\clearpage

\mbox{}
\begin{table}[h]
\begin{center}
\scriptsize
\begin{tabular}{||cc|c||cccc|c||ccc|c||cccc|c||cccc|c||}
\cline{4-22} \multicolumn{3}{c||}{} & \multicolumn{19}{c||}{} \\
\multicolumn{3}{c||}{} &
\multicolumn{19}{c||}{flavour channels and totals for $M(J\ell sn\; =\;0110)$}\\
\cline{4-22} \multicolumn{3}{c||}{} &
\multicolumn{14}{c||}{} & \multicolumn{5}{c||}{} \\
\multicolumn{3}{c||}{} &
\multicolumn{14}{c||}{$SU_{3}$-octet members} &
\multicolumn{5}{c||}{$SU_{3}$-singlets} \\
\hline \multicolumn{3}{||c||}{} &
\multicolumn{14}{c||}{} & \multicolumn{5}{c||}{} \\
\multicolumn{3}{||c||}{spatial q-numbers} & \multicolumn{5}{c||}{isotriplets} &
\multicolumn{4}{c||}{isodoublets} &
\multicolumn{5}{c||}{isoscalars} & \multicolumn{5}{c||}{} \\ [.3cm]
$M_{1}$ & $M_{2}$ & rel. & \multicolumn{5}{c||}{($t$)} &
\multicolumn{4}{c||}{($d$)} & \multicolumn{5}{c||}{($8$)} &
\multicolumn{5}{c||}{($1$)}\\ [.3cm]
$\!\! j\ell sn\!\!$ & $\!\! j\ell sn\!\!$ & $\!\! \ell sn\!\!$ &
$\!\! tt\!\!$ & $\!\! dd\!\!$ & $\!\! t8\!\!$ &
$\!\! t1\!\!$ & $\!\! T\!\!$ & $\!\! td\!\!$ & $\!\! d8\!\!$ &
$\!\! d1\!\!$ & $\!\! T\!\!$ & $\!\! tt\!\!$ & $\!\! dd\!\!$ &
$\!\! 88\!\!$ & $\!\! 11\!\!$ & $\!\! T\!\!$ & $\!\! tt\!\!$ &
$\!\! dd\!\!$ & $\!\! 88\!\!$ & $\!\! 11\!\!$ & $\!\! T\!\!$ \\
\hline & & & & & & & & & & & & & & & & & & & & & \\ [.3cm]
$\!\! 0000\!\!$ & $\!\! 0000\!\!$ & $\!\! 001\!\!$ &
- &
$\!\!\frac{    1}{   72}\!\!$ &
$\!\!\frac{    1}{  108}\!\!$ &
$\!\!\frac{    1}{   54}\!\!$ &
$\!\!\frac{    1}{   24}\!\!$ &
$\!\!\frac{    1}{   48}\!\!$ &
$\!\!\frac{    1}{  432}\!\!$ &
$\!\!\frac{    1}{   54}\!\!$ &
$\!\!\frac{    1}{   24}\!\!$ &
$\!\!\frac{    1}{   72}\!\!$ &
$\!\!\frac{    1}{  216}\!\!$ &
$\!\!\frac{    1}{  216}\!\!$ &
$\!\!\frac{    1}{   54}\!\!$ &
$\!\!\frac{    1}{   24}\!\!$ &
$\!\!\frac{    1}{   72}\!\!$ &
$\!\!\frac{    1}{   54}\!\!$ &
$\!\!\frac{    1}{  216}\!\!$ &
$\!\!\frac{    1}{  216}\!\!$ &
$\!\!\frac{    1}{   24}\!\!$\\ [.3cm]
$\!\! 0000\!\!$ & $\!\! 0001\!\!$ & $\!\! 000\!\!$ &
- &
$\!\!\frac{    1}{  144}\!\!$ &
$\!\!\frac{    1}{  216}\!\!$ &
$\!\!\frac{    1}{  108}\!\!$ &
$\!\!\frac{    1}{   48}\!\!$ &
$\!\!\frac{    1}{   96}\!\!$ &
$\!\!\frac{    1}{  864}\!\!$ &
$\!\!\frac{    1}{  108}\!\!$ &
$\!\!\frac{    1}{   48}\!\!$ &
$\!\!\frac{    1}{  144}\!\!$ &
$\!\!\frac{    1}{  432}\!\!$ &
$\!\!\frac{    1}{  432}\!\!$ &
$\!\!\frac{    1}{  108}\!\!$ &
$\!\!\frac{    1}{   48}\!\!$ &
$\!\!\frac{    1}{  144}\!\!$ &
$\!\!\frac{    1}{  108}\!\!$ &
$\!\!\frac{    1}{  432}\!\!$ &
$\!\!\frac{    1}{  432}\!\!$ &
$\!\!\frac{    1}{   48}\!\!$\\ [.3cm]
$\!\! 1010\!\!$ & $\!\! 1010\!\!$ & $\!\! 001\!\!$ &
- &
$\!\!\frac{    1}{  216}\!\!$ &
$\!\!\frac{    1}{  324}\!\!$ &
$\!\!\frac{    1}{  162}\!\!$ &
$\!\!\frac{    1}{   72}\!\!$ &
$\!\!\frac{    1}{  144}\!\!$ &
$\!\!\frac{    1}{ 1296}\!\!$ &
$\!\!\frac{    1}{  162}\!\!$ &
$\!\!\frac{    1}{   72}\!\!$ &
$\!\!\frac{    1}{  216}\!\!$ &
$\!\!\frac{    1}{  648}\!\!$ &
$\!\!\frac{    1}{  648}\!\!$ &
$\!\!\frac{    1}{  162}\!\!$ &
$\!\!\frac{    1}{   72}\!\!$ &
$\!\!\frac{    1}{  216}\!\!$ &
$\!\!\frac{    1}{  162}\!\!$ &
$\!\!\frac{    1}{  648}\!\!$ &
$\!\!\frac{    1}{  648}\!\!$ &
$\!\!\frac{    1}{   72}\!\!$\\ [.3cm]
$\!\! 1010\!\!$ & $\!\! 1011\!\!$ & $\!\! 000\!\!$ &
- &
$\!\!\frac{    1}{  432}\!\!$ &
$\!\!\frac{    1}{  648}\!\!$ &
$\!\!\frac{    1}{  324}\!\!$ &
$\!\!\frac{    1}{  144}\!\!$ &
$\!\!\frac{    1}{  288}\!\!$ &
$\!\!\frac{    1}{ 2592}\!\!$ &
$\!\!\frac{    1}{  324}\!\!$ &
$\!\!\frac{    1}{  144}\!\!$ &
$\!\!\frac{    1}{  432}\!\!$ &
$\!\!\frac{    1}{ 1296}\!\!$ &
$\!\!\frac{    1}{ 1296}\!\!$ &
$\!\!\frac{    1}{  324}\!\!$ &
$\!\!\frac{    1}{  144}\!\!$ &
$\!\!\frac{    1}{  432}\!\!$ &
$\!\!\frac{    1}{  324}\!\!$ &
$\!\!\frac{    1}{ 1296}\!\!$ &
$\!\!\frac{    1}{ 1296}\!\!$ &
$\!\!\frac{    1}{  144}\!\!$\\ [.3cm]
$\!\! 1010\!\!$ & $\!\! 1210\!\!$ & $\!\! 000\!\!$ &
- &
$\!\!\frac{    5}{  108}\!\!$ &
$\!\!\frac{    5}{  162}\!\!$ &
$\!\!\frac{    5}{   81}\!\!$ &
$\!\!\frac{    5}{   36}\!\!$ &
$\!\!\frac{    5}{   72}\!\!$ &
$\!\!\frac{    5}{  648}\!\!$ &
$\!\!\frac{    5}{   81}\!\!$ &
$\!\!\frac{    5}{   36}\!\!$ &
$\!\!\frac{    5}{  108}\!\!$ &
$\!\!\frac{    5}{  324}\!\!$ &
$\!\!\frac{    5}{  324}\!\!$ &
$\!\!\frac{    5}{   81}\!\!$ &
$\!\!\frac{    5}{   36}\!\!$ &
$\!\!\frac{    5}{  108}\!\!$ &
$\!\!\frac{    5}{   81}\!\!$ &
$\!\!\frac{    5}{  324}\!\!$ &
$\!\!\frac{    5}{  324}\!\!$ &
$\!\!\frac{    5}{   36}\!\!$\\ [.3cm]
$\!\! 1100\!\!$ & $\!\! 1100\!\!$ & $\!\! 000\!\!$ &
- &
$\!\!\frac{    1}{  144}\!\!$ &
$\!\!\frac{    1}{  216}\!\!$ &
$\!\!\frac{    1}{  108}\!\!$ &
$\!\!\frac{    1}{   48}\!\!$ &
$\!\!\frac{    1}{   96}\!\!$ &
$\!\!\frac{    1}{  864}\!\!$ &
$\!\!\frac{    1}{  108}\!\!$ &
$\!\!\frac{    1}{   48}\!\!$ &
$\!\!\frac{    1}{  144}\!\!$ &
$\!\!\frac{    1}{  432}\!\!$ &
$\!\!\frac{    1}{  432}\!\!$ &
$\!\!\frac{    1}{  108}\!\!$ &
$\!\!\frac{    1}{   48}\!\!$ &
$\!\!\frac{    1}{  144}\!\!$ &
$\!\!\frac{    1}{  108}\!\!$ &
$\!\!\frac{    1}{  432}\!\!$ &
$\!\!\frac{    1}{  432}\!\!$ &
$\!\!\frac{    1}{   48}\!\!$\\ [.3cm]
$\!\! 0110\!\!$ & $\!\! 0110\!\!$ & $\!\! 000\!\!$ &
- &
$\!\!\frac{    1}{   48}\!\!$ &
$\!\!\frac{    1}{   72}\!\!$ &
$\!\!\frac{    1}{   36}\!\!$ &
$\!\!\frac{    1}{   16}\!\!$ &
$\!\!\frac{    1}{   32}\!\!$ &
$\!\!\frac{    1}{  288}\!\!$ &
$\!\!\frac{    1}{   36}\!\!$ &
$\!\!\frac{    1}{   16}\!\!$ &
$\!\!\frac{    1}{   48}\!\!$ &
$\!\!\frac{    1}{  144}\!\!$ &
$\!\!\frac{    1}{  144}\!\!$ &
$\!\!\frac{    1}{   36}\!\!$ &
$\!\!\frac{    1}{   16}\!\!$ &
$\!\!\frac{    1}{   48}\!\!$ &
$\!\!\frac{    1}{   36}\!\!$ &
$\!\!\frac{    1}{  144}\!\!$ &
$\!\!\frac{    1}{  144}\!\!$ &
$\!\!\frac{    1}{   16}\!\!$\\ [.3cm]
$\!\! 1110\!\!$ & $\!\! 1110\!\!$ & $\!\! 000\!\!$ &
- &
$\!\!\frac{    1}{   36}\!\!$ &
$\!\!\frac{    1}{   54}\!\!$ &
$\!\!\frac{    1}{   27}\!\!$ &
$\!\!\frac{    1}{   12}\!\!$ &
$\!\!\frac{    1}{   24}\!\!$ &
$\!\!\frac{    1}{  216}\!\!$ &
$\!\!\frac{    1}{   27}\!\!$ &
$\!\!\frac{    1}{   12}\!\!$ &
$\!\!\frac{    1}{   36}\!\!$ &
$\!\!\frac{    1}{  108}\!\!$ &
$\!\!\frac{    1}{  108}\!\!$ &
$\!\!\frac{    1}{   27}\!\!$ &
$\!\!\frac{    1}{   12}\!\!$ &
$\!\!\frac{    1}{   36}\!\!$ &
$\!\!\frac{    1}{   27}\!\!$ &
$\!\!\frac{    1}{  108}\!\!$ &
$\!\!\frac{    1}{  108}\!\!$ &
$\!\!\frac{    1}{   12}\!\!$\\ [.3cm]
$\!\! 0000\!\!$ & $\!\! 1110\!\!$ & $\!\! 110\!\!$ &
- &
$\!\!\frac{    1}{   18}\!\!$ &
$\!\!\frac{    1}{   27}\!\!$ &
$\!\!\frac{    2}{   27}\!\!$ &
$\!\!\frac{    1}{    6}\!\!$ &
$\!\!\frac{    1}{   12}\!\!$ &
$\!\!\frac{    1}{  108}\!\!$ &
$\!\!\frac{    2}{   27}\!\!$ &
$\!\!\frac{    1}{    6}\!\!$ &
$\!\!\frac{    1}{   18}\!\!$ &
$\!\!\frac{    1}{   54}\!\!$ &
$\!\!\frac{    1}{   54}\!\!$ &
$\!\!\frac{    2}{   27}\!\!$ &
$\!\!\frac{    1}{    6}\!\!$ &
$\!\!\frac{    1}{   18}\!\!$ &
$\!\!\frac{    2}{   27}\!\!$ &
$\!\!\frac{    1}{   54}\!\!$ &
$\!\!\frac{    1}{   54}\!\!$ &
$\!\!\frac{    1}{    6}\!\!$\\ [.3cm]
$\!\! 1010\!\!$ & $\!\! 1100\!\!$ & $\!\! 110\!\!$ &
- &
$\!\!\frac{    1}{   18}\!\!$ &
$\!\!\frac{    1}{   27}\!\!$ &
$\!\!\frac{    2}{   27}\!\!$ &
$\!\!\frac{    1}{    6}\!\!$ &
$\!\!\frac{    1}{   12}\!\!$ &
$\!\!\frac{    1}{  108}\!\!$ &
$\!\!\frac{    2}{   27}\!\!$ &
$\!\!\frac{    1}{    6}\!\!$ &
$\!\!\frac{    1}{   18}\!\!$ &
$\!\!\frac{    1}{   54}\!\!$ &
$\!\!\frac{    1}{   54}\!\!$ &
$\!\!\frac{    2}{   27}\!\!$ &
$\!\!\frac{    1}{    6}\!\!$ &
$\!\!\frac{    1}{   18}\!\!$ &
$\!\!\frac{    2}{   27}\!\!$ &
$\!\!\frac{    1}{   54}\!\!$ &
$\!\!\frac{    1}{   54}\!\!$ &
$\!\!\frac{    1}{    6}\!\!$\\ [.3cm]
$\!\! 1010\!\!$ & $\!\! 1010\!\!$ & $\!\! 220\!\!$ &
- &
$\!\!\frac{    5}{   54}\!\!$ &
$\!\!\frac{    5}{   81}\!\!$ &
$\!\!\frac{   10}{   81}\!\!$ &
$\!\!\frac{    5}{   18}\!\!$ &
$\!\!\frac{    5}{   36}\!\!$ &
$\!\!\frac{    5}{  324}\!\!$ &
$\!\!\frac{   10}{   81}\!\!$ &
$\!\!\frac{    5}{   18}\!\!$ &
$\!\!\frac{    5}{   54}\!\!$ &
$\!\!\frac{    5}{  162}\!\!$ &
$\!\!\frac{    5}{  162}\!\!$ &
$\!\!\frac{   10}{   81}\!\!$ &
$\!\!\frac{    5}{   18}\!\!$ &
$\!\!\frac{    5}{   54}\!\!$ &
$\!\!\frac{   10}{   81}\!\!$ &
$\!\!\frac{    5}{  162}\!\!$ &
$\!\!\frac{    5}{  162}\!\!$ &
$\!\!\frac{    5}{   18}\!\!$
 \\ & & & & & & & & & & & & & & & & & & & & & \\ \hline
\end{tabular}
\normalsize
\end{center}
\caption[]{Coupling constants for the decay processes of scalar
mesons into meson pairs. The interpretation of the content of the table is
explained in the text.}
\label{scalar}
\end{table}
\clearpage

\begin{table}[h]
\begin{center}
\begin{tabular}{|l||l|l|}
\hline & & \\
initial meson & \multicolumn{2}{c|}{decay products} \\ [.3cm]
 & \multicolumn{1}{c|}{\cite{scalar}}
 & \multicolumn{1}{c|}{\cite{Toern95}} \\ & & \\ \hline & & \\
$a_{0}$ or $\delta$ & 
$\fnd{1}{3}$ ($K\bar{K}$) \,+\, $\fnd{2}{3}$ ($\pi\eta_n$) &
$\fnd{1}{3}$ ($K\bar{K}$) and $\fnd{2}{3}$ (sum $\pi\eta$'s) \\ [.3cm]
$K^{\ast}_{0}$ or $\kappa$ &
$\fnd{1}{2}$ ($K\pi$) \,+\, $\fnd{1}{2}$ ($K\eta_n$+$K\eta_s$) &
$\fnd{1}{2}$ ($K\pi$) and $\fnd{1}{2}$ (sum $K\eta$'s) \\ [.3cm]
$f_{0}$ or $\epsilon /S$ $n\bar{n}$ &
$\fnd{3}{5}$ ($\pi\pi$) \,+\, $\fnd{1}{5}$ ($K\bar{K}$) \,+\, $\fnd{1}{5}$
($\eta_n\eta_n$) &
1 ($\pi\pi$), $\fnd{1}{3}$ ($K\bar{K}$) and $\fnd{1}{3}$
(sum $\eta\eta$'s) \\ [.3cm]
$f_{0}$ or $\epsilon /S$ $s\bar{s}$ &
$\fnd{1}{2}$ ($K\bar{K}$) \,+\, $\fnd{1}{2}$ ($\eta_s\eta_s$) &
$\fnd{2}{3}$ ($K\bar{K}$) and $\fnd{2}{3}$ (sum $\eta\eta$'s)\\
 & & \\ \hline
\end{tabular}
\end{center}
\caption[]{Quadratic coupling constants for the decay process of a scalar
meson into a pair of pseudoscalar mesons.}
\label{tabspsps}
\end{table}
\end{document}